\documentclass[aps,pra,superscriptaddress,twocolumn,amsmath,amssymb,onecolumn]{revtex4}
\usepackage{bbm}
\usepackage{graphicx}

\newcommand{\Op}[1]{\boldsymbol{\mathsf{\hat{#1}}}}

\newcommand{\Fkt}[1]{\,\mathsf {#1}}

\def\eps{\varepsilon}

\begin{document}

\title{Stabilization of Ultracold Molecules Using Optimal Control Theory}
\date{\today}
\author{Christiane P. Koch}
\email{christiane.koch@lac.u-psud.fr}
\affiliation{Laboratoire Aim\'e Cotton, CNRS, B\^{a}t. 505, Campus d' Orsay,
91405 Orsay Cedex, France}
\affiliation{Department of Physical Chemistry and
The Fritz Haber Research Center, 
The Hebrew University, Jerusalem 91904, Israel}
\author{Jos\'{e} P. Palao}
\affiliation{Department of Physical Chemistry and
The Fritz Haber Research Center, 
The Hebrew University, Jerusalem 91904, Israel}
\affiliation{Departamento de F\'{i}sica Fundamental II,
Universidad de La Laguna, La Laguna 38204, Spain}
\author{Ronnie Kosloff}
\affiliation{Department of Physical Chemistry and
The Fritz Haber Research Center, 
The Hebrew University, Jerusalem 91904, Israel}
\author{Fran\c{c}oise Masnou-Seeuws}
\affiliation{Laboratoire Aim\'e Cotton, CNRS, B\^{a}t. 505, Campus d' Orsay,
91405 Orsay Cedex, France}

\begin{abstract}
In recent experiments on ultracold matter, 
molecules have been produced from ultracold atoms 
by photoassociation, Feshbach resonances, and three-body recombination. 
The created molecules are translationally cold, but 
vibrationally highly excited. This will eventually lead them to be lost 
from the trap due to collisions. We  propose
shaped laser pulses to transfer these highly excited 
molecules to their ground vibrational level. Optimal control
theory is employed to  find the light
field that will carry out this task with minimum intensity.
We present results for the sodium dimer. The final target can be
reached to within 99\% if the initial guess field is physically motivated.
We  find that the optimal fields  contain the transition frequencies
required by a  good Franck-Condon pumping scheme.
The analysis is able to identify the ranges of intensity and 
pulse duration which are able to achieve this task before other competing process 
take place. Such a scheme could produce stable ultracold molecular samples or 
even stable molecular Bose-Einstein condensates.
\end{abstract}

\maketitle

\section{Introduction}
\label{sec:intro}

The creation of cold molecules from atomic Bose-Einstein condensates 
(BEC)~\cite{DonleyNat02,RegalNat03,HerbigSci03,DuerrPRL04,XuPRL03,JochimSci03,GreinerNat03,ZwierleinPRL03} 
as well as from ultracold thermal gases~\cite{NicoPRL02,ChinPRL03}
has advanced remarkably over the last two years.
In both cases molecules are formed due to interaction of atoms
with an external field. The latter can be 
the electric field of a laser leading to photoassociation~\cite{FrancoiseReview} 
 or a magnetic field tuned to drive the atoms across
 a Feshbach resonance~\cite{Timmermans99}. 
Tuning  close to a Feshbach resonance can  furthermore be used to obtain
a very large scattering length.
This enhances three-body recombination and  molecules are thereby
formed~\cite{JochimSci03}.
In most experimental schemes
 translationally cold, but vibrationally very highly excited molecules are produced.
These highly excited molecules are not stable with respect to 
collisions, and dimers consisting of bosonic atoms in particular are very
rapidly. In the case of sodium this happens within milliseconds~\cite{XuPRL03}.
Creation is therefore only a first step toward novel
experiments using ultracold molecules~\cite{MeijerReview}, 
and their stabilization is the obvious next task. 

The ultimate goal is to create $v=0,J=0$ molecules. 
The task of transferring the highly excited molecules
to $v=0, J=0$ is far from being trivial: a long-range molecule and
a $v=0$ molecule are very different (Cf. Fig.~\ref{fig:scheme} where
the wavefunctions for 
the last bound and the ground vibrational level of the ground state of Na$_2$
are drawn).
This task has to be solved under three major constraints:
1.) The goal has
to be achieved in a time short compared to the collisional lifetime. 
The rate for collisional decay is not precisely known at present, but lifetimes 
on the order of or shorter than ms should be expected~\cite{XuPRL03}.
Calculations for collisional decay of molecules in highly excited vibrational
levels have been performed in the non-reactive $^{3,4}$He + H$_2(v)$ 
case~\cite{BalakrishnanPRL98}, showing an increase of the relaxation rate by
nearly three orders of magnitude  when the initial state is going from $v$=1 (rate 
$\sim$ 5$\times$ 10$^{-17}$~cm$^3$ s$^{-1}$) to $v$=10. In the reactive 
Na + Na$_2(v)$ case, quantum 
calculations for the $v=1\rightarrow 0$  rate yield  
$\sim 5\times 10^{-10}$cm$^3$ s$^{-1}$~\cite{SoldanPRL02}.
Therefore, at an atomic  density of 10$^{11}$ cm$^{-3}$, a vibrational 
quenching relaxation time well below 1 ms is indeed to be expected.
2.) If laser pulses are to be used, spontaneous emission has to be avoided. 
 The radiative lifetime of the excited state is on the order of 10~ns.
3.) Due to vibrational energy "pooling" or
vibration-to-vibration ladder climbing~\cite{TreanorJCP67}
the intermediate range of
binding energies needs to be avoided at any cost, i.e. the molecules 
need to be immediately transferred to the lowest levels.

In this paper, we suggest to employ optimal control to transfer
the highly excited molecules to the rovibrational ground state. 
Optimal control has been intensely studied 
both theoretically and experimentally in many areas of physical chemistry. 
Optimal control theory (OCT)~\cite{RiceBook} offers the prospect of driving an
atomic or molecular system
to an arbitrary, desired state due to the interaction with an external field. 
Experimentally, control is achieved via feedback learning loops, see e.g. 
\cite{BrixnerCPC03} for a recent review.
The difference between the approach we suggest and control experiments
as they have been performed over the last decade lies in the different target:
In the latter the goal consists usually in varying the ratio between
different dissociation channels, i.e. many final states are available 
and the target is rather broad, while
in the present context a single ro-vibrational level is to be achieved.
However, this does not pose a problem in principle -- it might
make control harder to achieve, but it does not render it impossible. 
We suggest optimal control to transfer the highly excited 
molecules to the rovibrational ground state
because it provides an {\em efficient} method operating
on a very {\em short} timescale. 

Alternatively, two-photon Raman transitions have been suggested~\cite{HerbigSci03} 
to transfer the 
highly excited molecules to low-lying vibrational states.  
However, if really the lowest vibrational levels shall be populated,
a Raman scheme with continuous wave (CW) laser pulses is not conceivable due to 
unfavorable Franck-Condon overlaps. The framework of 
\textbf{Sti}mulated \textbf{R}aman \textbf{A}diabatic \textbf{P}assage 
(STIRAP) is not any more promising since a highly oscillatory wave function with 
80\% or more of its probability density 
in the asymptotic region can not be transferred 
adiabatically into a very compact wave function, typically of Gaussian shape
(Cf. Fig.~\ref{fig:scheme}). 
This has been overlooked in previous 
studies~\cite{MackiePRL00,DrummondPRA02}
in which the vibrational degree of freedom was not explicitly taken into 
account. 

Optimal control theory has been applied before in the context of the formation of
ultracold molecules, specifically to calculate optimal conversion of 
an atomic into a molecular BEC~\cite{HornungPRA02b}. In this work, a laser
pulse was employed together with a magnetic field which induced a 
Feshbach resonance. The authors worked in the framework of
the Gross-Pitaevski equation which is a {\em non-linear} Schr\"odinger equation
while employing a {\em linear} variant of OCT. 
A nonlinear version of OCT as required for the application
to the Gross-Pitaevski equation has been worked out recently~\cite{SklarzPRA02}.
The vibrational structure of the molecules was simply 
taken into account by assigning a few vibrational levels in 
Ref.~\cite{HornungPRA02b}, but the 
coordinate-dependence of the vibrational wave functions, in particular
the qualitatively different character of initial and final state
wave functions, was neglected. Finally, the initial state in 
Ref.~\cite{HornungPRA02b} corresponds to two free, colliding atoms.
If one was to include the vibrational structure in the model of
Ref.~\cite{HornungPRA02b}, the resulting Hilbert space would be 
infinite-dimensional, and due to the theorem of 
controllability~\cite{RamakrishnaPRA95,TarnClark}
one would not be guaranteed that an optimal solution exists at all.

For conceptual clarity, we will therefore separate the creation of 
molecules, however weakly bound, from the process of their stabilization. 
This approach  corresponds furthermore nicely to the current 
status of experiments on cold molecules which create weakly bound
dimers via magnetic Feshbach resonances or three-body 
recombination~\cite{RegalNat03,HerbigSci03,DuerrPRL04,XuPRL03,JochimSci03,GreinerNat03,ZwierleinPRL03}. 
The stabilization of these
molecules in the experiment still remains an open problem to which 
no easy answers exist.
In this paper, we will hence start from extremely weakly bound molecules
and employ optimal control theory
to obtain short, shaped laser pulses which drive the system from a specified,
highly excited vibrational level to the ground state. 

We will apply the optimal control algorithm to the formation of {\em stable} 
sodium dimers. From molecular 
spectroscopy experiments with CW lasers~\cite{ElbsPRA99}, 
we know that a pathway connecting the rovibrational ground state
with the last bound levels exists.
Furthermore, sodium has been one of the first systems to be studied by femtosecond
 spectroscopy~\cite{BaumertPRL90} and control experiments~\cite{ShnitmanPRL96}.
Its properties under ultracold conditions are equally well 
studied, see e.g.~\cite{McKenziePRL02,XuPRL03}.
The scheme we suggest is similar in spirit
to that of Ref.~\cite{VardiJCP97} where nanosecond pulses were used.
In order to compare to current experimental
control techniques, however,  we employ pulses of femtosecond and
picosecond timescale. These pulses are optimized by the control algorithm
while they would serve as input for a feedback learning loop 
in a prospective experiment, see e.~g.~\cite{BrixnerCPC03}. This is 
 in contrast to the intuitively chosen pulses of Ref.~\cite{VardiJCP97}.
A large number of theoretical studies on femtosecond pulse control of
sodium has been reported, mostly in the context of photodissociation and
ionization, see e.g.~\cite{MachholmJCP96,PaciJCP98,ShenCPL03}. Our approach has
been motivated by the experiment of Ref.~\cite{ElbsPRA99} where the last bound
levels of the Na$_2$ ground state were probed. This was achieved by exciting
$v=0$ Na$_2$ molecules from a molecular beam with CW lasers
to an excited electronic state and subsequent spontaneous emission. 
A scheme with multiple excitation-deexcitation steps was necessary 
due to Franck-Condon overlaps. 
In principle, one could  invert 
this scheme to create $v=0$ from highly excited molecules.
However, the yield would be low and the required times
cannot compete with the time scales of collisional loss and spontaneous
emission. We therefore look for optimal 
ultrashort pulses which realize this inverse scheme.

We formulate the problem as one of optimizing state-to-state transitions. This is 
in contrast to a density matrix formulation which describes transitions from an
ensemble of states to another one~\cite{AllonJCP93,JosePRL02}.  
As an extreme case, the latter includes 
true cooling as a transition from an ensemble of states to a single state. 
Note, that true cooling requires coupling 
to a dissipative environment which accepts
the entropy~\cite{AllonJCP97,AllonCP01}.
However, the optimization of state-to-state transitions should 
be sufficient for Bose-condensed samples since one starts with a coherent sample. 
Even in the case of
thermal samples state-to-state transitions can serve as an important first step
in which  one state is selected to be transferred while  all others are
ionized. We will employ OCT in the Krotov variant
restricting the change in pulse energy at each 
iteration~\cite{SklarzPRA02,AllonCP01,JosePRA03}. 
This guarantees monotonous and smooth
convergence toward the objective.

We will work in the framework of the linear Schr\"odinger equation which describes
the internuclear dynamics of two atoms, i.e. in case of a 
Bose-Einstein condensate (BEC) we neglect 
the condensate dynamics. This approach is justified by the time scales present
in standard optical and/or magnetic traps.
While the internuclear dynamics and pulse shaping occur on the time scale of 
femtoseconds to picoseconds, the condensate dynamics for conventional traps is
characterized by microseconds. The condensate will have to adjust to the 
new internal state, but  this is going to happen on a much slower time scale
than the one of the pulse~\cite{PascalPRA03}.
We can hence safely ignore the influence of the condensate
on the dynamics of the state-to-state transition and assume that 
the adjustment of the
condensate takes place once the pulse is over.

The paper is organized as follows: 
The model for
the sodium dimer as well as the proposed scheme to form stable ultracold molecules
are described in Section~\ref{sec:model}.
Section~\ref{sec:opt} briefly reviews the 
optimal control algorithm with details given 
in the Appendix.  The results are presented in
 Section~\ref{sec:results}. Section~\ref{sec:concl} concludes with a discussion of
the experimental feasibility of our proposed scheme as well as its implications
to vibrational cooling.

\section{The Na$_2$ system}
\label{sec:model}

We have chosen Na$_2$ because this system
has been intensely studied in ultracold experiments, in 
control experiments, and by traditional spectroscopy, 
and a large amount of experimental data is 
available. In particular, highly accurate potential energy curves
have been obtained from molecular spectroscopy~\cite{SamuelisPRA00,TiemannZfPD96}.
Experiments~\cite{ElbsPRA99} have furthermore
shown that a route from the last bound levels to $v=0$ exists.
In this scheme, ground state molecules from a molecular beam
with $v=0$, $J=0$ are excited by a CW laser 
($\lambda=610$~nm) to the A$^1\Sigma^+_u$ excited state ($v^\prime=15$). Those
molecules which decay to the $v=29$ level of the ground state, are excited
by a second CW laser ($\lambda=540$~nm) to excited state levels with 
$v^\prime=100-140$. A third CW laser ($\lambda=595$~nm) probes
the transition between these excited state levels and the last bound levels
($v=61-65$) of the ground state. 

\subsection{Proposed scheme for the formation of ultracold, stable molecules}
\label{subsec:scheme}
\begin{figure}[tb]
  \centering
  \includegraphics[width=0.9\linewidth]{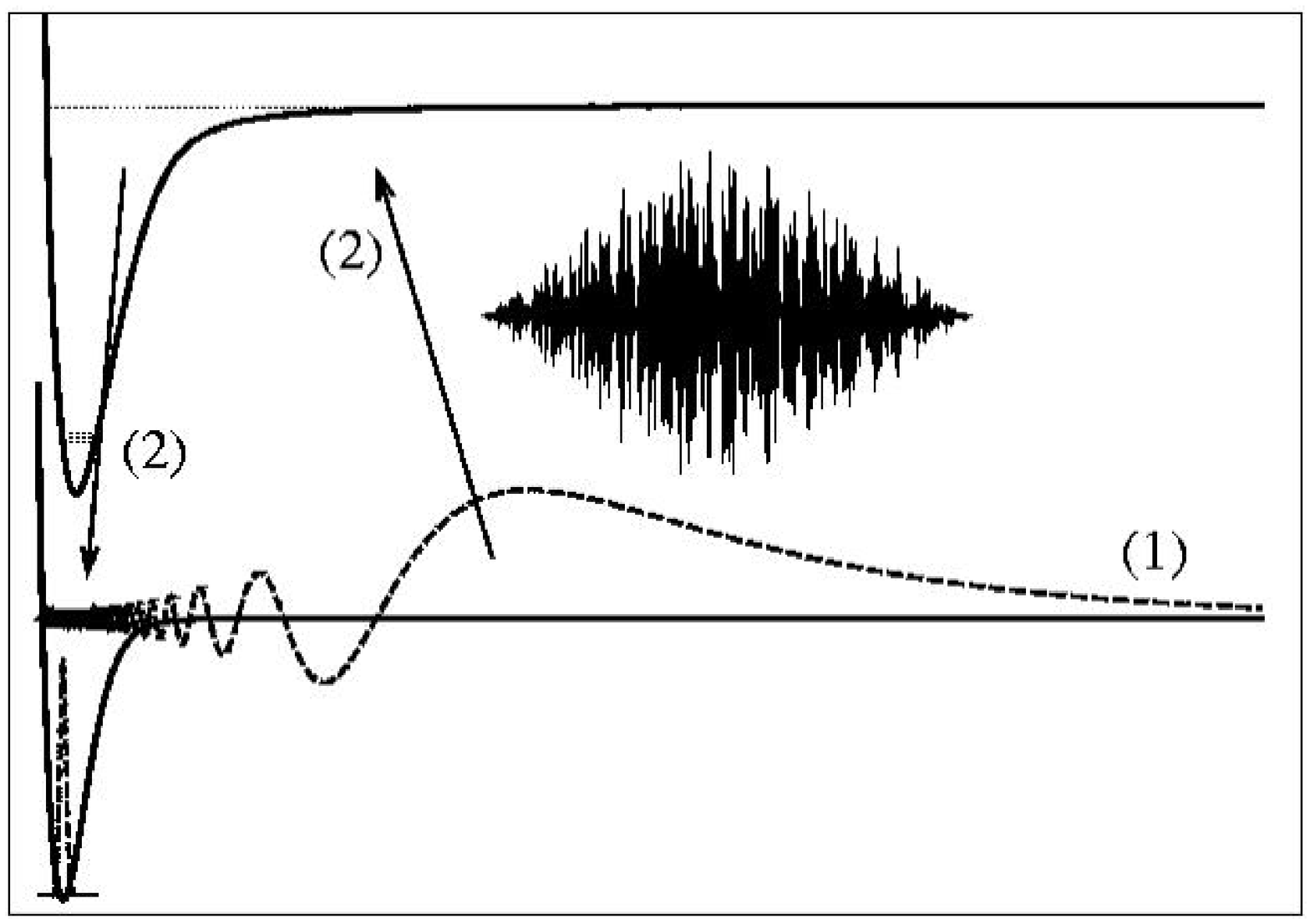}
  \caption{Scheme for the production of stable ultracold molecules:
    Vibrationally highly excited  molecules are created due to the interaction
    with an external field (1) and are transferred to the ground state by
    a shaped laser pulse (2). The wave functions of the 
    last bound and the $v=0$ levels are drawn as dashed lines, while the 
    excited state vibrational levels which have good Franck-Condon overlap
    with the last level and with the ground level, respectively,
    are indicated by dotted lines. 
    Obviously, no direct transition exists.
  }
  \label{fig:scheme}
\end{figure}

We envisage the following two-step scheme for the production of ultracold 
molecules (Fig.~\ref{fig:scheme}).
First, loosely bound molecules are created by tuning an external
magnetic field to sweep an atomic sample across a Feshbach 
resonance~\cite{RegalNat03,HerbigSci03,DuerrPRL04,XuPRL03}
or by enhanced three-body recombination in the vicinity of a Feshbach
resonance~\cite{JochimSci03}. Another
possibility is given by applying a weak off-resonant CW photoassociation 
laser. The last level is so extremely loosely bound that even a very weak
field perturbs it leading to a coupling with the continuum 
states~\cite{Philpel}. 
Second, a shaped laser pulse is applied to
transfer the highly excited molecules to $v=0$. This second step
is of our concern in the present study.
Following the experiment~\cite{ElbsPRA99}, we propose 
to employ  fields which induce transitions between the 
Na$_2$ ground state ($X^1\Sigma^+_g$ ) and the A$^1\Sigma^+_u$ excited state.

One could imagine a one-step scheme where the optimal pulse is used to 
directly transfer a continuum wave function to the 
ro-vibrational ground state. 
However, in the case
of continuum states the theorem of controllability~\cite{RamakrishnaPRA95,TarnClark}
 is not applicable and one is not guaranteed that an optimal solution 
exists. Furthermore, the results may depend on how
the continuum is modelled. Since experimentally, step (1) seems to
become a standard technique~\cite{RegalNat03,HerbigSci03,DuerrPRL04,XuPRL03}, 
we prefer to restrict the task of the optimal pulse to the transfer of
one bound level to another bound level and to attain a 
mathematically well-defined model.

\subsection{Model and methods}
\label{subsec:methods}

We are interested in obtaining an estimate of the feasibility of optimal
control experiments with ultracold molecules. We therefore restrict our approach
to a qualitative model with two channels which correspond to the two 
electronic states of Ref.~\cite{ElbsPRA99}, and we
neglect hyperfine interaction, spin-orbit coupling 
and rotational excitation. These effects should be included
to make {\em quantitative} predictions.

Taking hyperfine interaction into account would
amount to studying a model with more than  two channels. 
At the binding energies at which cold molecules are created in the 
current experiments, a single channel description for the ground state
is not valid~\cite{SamuelisPRA00}, 
i.e. the wave function contains singlet as well as
triplet components. The shaped laser pulse does not alter the coherence
of this superposition. A multi-channel treatment would therefore result
in a pulse with parts corresponding to the excitation pathway of singlet,
and other parts corresponding to the excitation pathway of triplet 
molecules. However, in this study
we want to emphasize the importance of the internal structure of
the molecules, i.e. the coordinate-dependence of the atom-atom
interaction potentials, for the creation of {\em stable} molecules.
We therefore restrict the model to two channels, one for the ground and one
for the electronically excited state, and we perform calculations for 
singlet states. This amounts to neglecting the triplet component of the
wave function. If the triplet component of the wave function is 
larger than the singlet component, one would have to repeat the 
present calculations using the corresponding triplet potentials. The
qualitative principle, however, remains unaltered. 

The argument concerning rotational excitation
is along the same lines as in the case of hyperfine 
interaction. Allowing for rotational excitation would
amount to treating additional channels with the centrifugal barrier 
for each $J$ added to the atom-atom interaction.
The additional channels increase the size of the search space, but at the
same time allow for more pathways, i.e. for more flexibility. 
It is thus not clear {\em a priori} whether the search for optimal
pulses will be slowed down or sped up.
Moreover, if the initial state contains some rotational excitation, 
one might speculate that this facilitates control since such a
wave function would be more bound than one with $J=0$. 
Of course, such a claim can only be proven by 
a detailed investigation which is beyond the scope of the present work.

We  solve the radial Schr\"odinger equation for the 
vibrational motion of the sodium dimer,
\begin{equation}
  \label{eq:schroed}
  i\hbar \,\frac{\partial}{\partial t}\,\varphi(R,t) = \Op{H}\, \varphi(R,t) \,,
\end{equation}
with $R$ the distance between the two nuclei. 
The vibrational wave function consists of two components,
$\varphi=\begin{pmatrix} \varphi_g \\ \varphi_e \end{pmatrix}$,
corresponding to the two channels.
The Hamiltonian describing two electronic states and the nuclear 
degree of freedom, $R$, is given by
\begin{equation}
  \label{eq:H}
  \Op{H} =  \begin{pmatrix}
    \Op{T} + \Op{V}_g  & 0 \\ 
    0 &  \Op{T}+\Op{V}_e
  \end{pmatrix} +
  \begin{pmatrix}
    0 & \Op{\mu}\, \varepsilon(t) \\
    \Op{\mu}\, \varepsilon(t)^* & 0
  \end{pmatrix} \,,
\end{equation}
where $\Op{T}=\frac{\hbar}{2m}\frac{\partial}{\partial R}$ denotes 
the kinetic energy operator, 
$\Op{\mu}$ the dipole operator and $\varepsilon(t)$ the electric field.
The ground state potential $\Op{V}_g$ describes the state of $X^1\Sigma_g$ 
symmetry which is correlated to the (3s+3s) asymptote of sodium, and the 
excited state potential $\Op{V}_e$ describes the state of $A^1\Sigma_u$ 
symmetry correlated to the (3s+3p) asymptote. The ground state potential has
been obtained by an analytical fit to high-resolution spectroscopic 
data~\cite{SamuelisPRA00}, while an RKR potential~\cite{TiemannZfPD96} has 
been matched to an asymptotic expansion~\cite{MarinescuPRA95} 
for the excited state potential.
We assume the dipole operator to be independent of $R$, $\Op{\mu}=\,$const. While
this approximation needs to be improved, it might well serve as a first step
since the two concerned electronic states do not show avoided crossings.
The electric field is taken to be real which means that a fixed 
(but not specified) polarization is assumed. The initial guess field is 
given by $\varepsilon(t) = \varepsilon_0 \,S(t) \cos(w t)$ where $S(t)$ is a Gaussian
envelope function or a sequence of Gaussians 
and $\omega$ is the central frequency of the pulse.

The wave function $\varphi(R,t)$
is represented on a grid employing the mapped Fourier grid
method~\cite{SlavaJCP99} which allows for accurately representing even the 
last bound levels of the ground state potential 
with a comparatively small number of grid points ($N=1024$). The mapped 
Fourier grid method introduces, however, 
unphysical states which can lead to spurious
effects in the dynamics. For the sodium system, these "ghost" states can
be eliminated by using a basis of sine functions (instead of plane 
waves)~\cite{WillnerJCP04}
and by choosing the density of points sufficiently high. 
The mapped grid is calculated from the envelope potential of both
$\Op{V}_g$ and $ \Op{V}_e$, i.e. the same grid is used on both channels.

The time-dependent Schr\"odinger equation, Eq.~(\ref{eq:schroed}), is solved
formally by
\begin{equation}
  \label{eq:prop}
  \varphi(R,t) = \Op{U}(t,0)\, \varphi(R,0) \,. 
\end{equation}
We employ a Chebychev expansion~\cite{RonnieReview94,RonnieGrid96}
 of the time evolution operator, $\Op{U}(t)$,
which is numerically exact for a time-independent Hamiltonian. In our case
of an explicitly time-dependent Hamiltonian, Eq.~(\ref{eq:prop}) is second 
order in the time step. Note, that we cannot use a less expensive propagation 
method such as the split propagator since the grid mapping introduces 
multiplications between functions in Fourier space and functions in real 
space, i.e. terms of the kind $f(R) \frac{\partial}{\partial R}$.

The initial state as well as the target state are taken to be eigenstates of 
the ground state potential. The eigenstates are computed simply by 
calculating a matrix representation of the Hamiltonian in the sine basis
and subsequent diagonalization~\cite{WillnerJCP04}.

\section{The Optimal Control Algorithm}
\label{sec:opt}
We formulate the optimal control for state-to-state transitions, i.e. we want to find
the field which drives a specified initial state $|\varphi_i\rangle$
to a specified target state $|\varphi_f\rangle$ at the final time $t=T$. The
objective we want to reach can therefore be defined as the overlap between 
the initial state  propagated 
from time $t=0$ to $t=T$ with field $\varepsilon(t)$ and  the target
state,
\begin{equation}
  \label{eq:goal}
  F = |\langle \varphi_i|\Op{U}^+(T,0;\varepsilon)
  |\varphi_f\rangle|^2 \,.
\end{equation}
$\Op{U}^+(T,0;\varepsilon)$ denotes the evolution operator which
completely specifies the system dynamics. In the calculations presented below,
$|\varphi_i\rangle$ will be a highly excited vibrational level
of the Na$_2$ ground state, while $|\varphi_f\rangle$ will be chosen 
to be the vibrational ground state of the Na$_2$ ground state.  The length
of the optimization time interval, $T$, is an external parameter of the 
calculation. If one compares to experiment, it is related to the
bandwidth of the pulse and the number of pixels of the pulse shaper. 
Taking $T$ to be
twice the longest vibrational period which occurs in the problem is usually a 
good guess. In the case of long-range molecules, 
very different time scales are involved which  
leads to a problem of feasibility. We present a work-around
 in Sec.~\ref{subsec:T}. A field is optimal if it almost completely
transfers the initial state $|\varphi_i\rangle$ to the target $|\varphi_f\rangle$,
i.e. if $F$ attains a value close to one.

The objective $F$ is a functional of the field $\varepsilon(t)$. However, it
depends explicitly only on the final time $T$. To use information from
the dynamics at intermediate times, i.e. within the time interval $(0,T)$, 
we define a new functional $J$, 
\begin{equation}
  \label{eq:J}
  J = -F + \int_0^T g(\varepsilon,\varphi) \mathrm{d}t\,,
\end{equation}
where the integral term denotes additional constraints over the system evolution. 
The optimal field will be found by minimization of $J$. The additional constraints
provide the connection between the dynamics and reaching the objective, they
therefore define how the control is accomplished~\cite{SklarzPRA02}. 
This becomes
particularly apparent in the Krotov variant of optimal control theory
(see~\cite{SklarzPRA02} for a review) which we will employ in the 
following. The Krotov method allows us to derive an iterative procedure which
maximizes the original functional $F$ (minimizes $-F$) 
at the final time $T$ while
modifying the field at intermediate times and guaranteeing monotonic 
convergence. Details can be found in Appendix~\ref{sec:krotov}.

The choice of $g(\varepsilon,\varphi)$ is not arbitrary, 
it needs to comply with
the requirement of monotonic convergence~\cite{JosePRA03}
(see also Appendix~\ref{sec:krotov}). In this 
study, we use
\begin{equation}
  \label{eq:g}
  g(\varepsilon,\varphi) = g(\varepsilon) = \frac{\alpha}{S(t)} 
  \left[\varepsilon(t)-\tilde{\varepsilon}(t) \right]^2  \,,
\end{equation}
where $\tilde{\varepsilon}(t)$ is the field of the previous iteration.
$S(t)$ denotes the constraint to smoothly switch the
field on and off with  shape function $S(t)$. We take it to be 
Gaussian or $S(t)=\sin^2(\pi t/T)$. 
Eq.(\ref{eq:g}) has the physical interpretation of restricting the change in
pulse energy at each iteration. It allows for a smooth convergence toward the
objective since the change in field vanishes when the optimal field is
approached~\cite{JosePRA03}. The optimization strategy can be
controlled by the parameter $\alpha$: A small value results in a small
weight of the additional constraint and allows for large modifications in the
field, while a large value of $\alpha$ represents a conservative search
strategy allowing only small modifications of the field at each iteration.
The parameter $\alpha$ is not directly related to the number of 
iterations required to find a solution, and the best search strategy
is usually given by running the optimal control algorithm first
with a small and then with a large value of $\alpha$~\cite{HornungPRA02a}:
A small $\alpha$ will excite many pathways, then the solution
can be refined with large $\alpha$. 

With this choice of $g(\varepsilon,\varphi)$ the change in field,
$ \Delta \varepsilon = \varepsilon^\mathrm{new} - \varepsilon^\mathrm{old}$,
at every time $t$, $0<t<T$, is given by 
\begin{eqnarray}
  \label{eq:deltaeps}
  \Delta \varepsilon(t) &=& \frac{S(t)}{\alpha} \;\mathfrak{Im} \left[\;
    \langle \underbrace{\varphi_i|\,\Op{U}^+(T,0;\varepsilon^\mathrm{old})}
    \,|\varphi_f \rangle\;
    \langle  \underbrace{\varphi_f|
      \,\Op{U}^+(t,T;\varepsilon^\mathrm{old})}\, \Op{\mu}\,  
    \underbrace{\Op{U}(t,0;\varepsilon^\mathrm{new})\,|
      \varphi_i\rangle }\;   \right] \\
    &&  \quad \quad \quad  \quad \quad \quad \quad
    \mathrm{forward} \quad \quad  \quad \quad \quad 
    \mathrm{backward}  \quad \quad \quad \quad
    \mathrm{forward} \nonumber
\end{eqnarray}
(see Appendix~\ref{sec:krotov} for the derivation).
The change in field is determined by the "expectation value" of the dipole
operator $\Op{\mu}$, 
i.e. by matching at time $t$ the initial wave 
function which has been propagated forward in time from $t=0$ to $t$
with the target wave function
which has been propagated backward in time from $T$ to $t$. 
Note, that this is not an ordinary quantum mechanical expectation value
due to the different fields, $\varepsilon^\mathrm{old}$ of the previous
iteration and $\varepsilon^\mathrm{new}$ of the current iteration. 
We consider two channels/electronic states for the Na$_2$ system.
The wave function therefore contains two components,
$\begin{pmatrix} \varphi_g \\ \varphi_e \end{pmatrix}$, and 
the dipole operator, $\begin{pmatrix} 0 & \Op{\mu} \\ \Op{\mu} & 0\end{pmatrix}$,
induces electronic transitions. Note that while for the initial and target state
$\varphi_e =0$, the propagated states have a non-zero  $\varphi_e$-component 
due to the field.

Eq.~(\ref{eq:deltaeps}) implies that 
information from the dynamics in both the past and the future is used to 
calculate the new field. The iterative algorithm is thus defined:
We first pick a guess field $\varepsilon^\mathrm{old}$, propagate $|\varphi_i\rangle$ 
with $\Op{U}(T,0;\varepsilon^\mathrm{old})$ (forward) and evaluate the objective. We
then propagate the target state $|\varphi_f\rangle$ from time $t=T$ to
time $t=0$ (backward), storing the propagated wave function for all intermediate
times. In a third step we obtain the new field $\varepsilon^\mathrm{new}$ at each 
time $t$ by evaluating
Eq.~(\ref{eq:deltaeps}) and we use $\varepsilon^\mathrm{new}$ 
to propagate $|\varphi_i\rangle$ 
(forward). These steps have to be repeated until the objective
has reached the desired value close to one.
 
Note that
since $\Delta \varepsilon =\varepsilon^\mathrm{new}-\varepsilon^\mathrm{old} $, 
Eq.~(\ref{eq:deltaeps}) is 
implicit in the new field $\varepsilon^\mathrm{new}$.
 A simple, but sufficient remedy to
avoid  solving this implicit equation
is found by employing two different grids in the time discretization~\cite{JosePRA03}.
The time grid for the wave functions
 has $N_t+1$ points and is defined from $t=0$ to $t=T$, while the 
grid where the field is evaluated has $N_t$ points and ranges
from $t=\Delta t/2$ to $t=T - \Delta t/2$ with $\Delta t = T/N_t$ the spacing
for both grids. The new field $\varepsilon^\mathrm{new}$ 
at the first interleaved grid point $t=\Delta t/2$ 
is then calculated from the wave function at $t=0$, i.e. from
$\Op{U}(0,0;\varepsilon^\mathrm{new})\,| \varphi_i\rangle =| \varphi_i\rangle $, 
while the wave function is propagated from
$t=0$ to $t=\Delta t$ using the field $\varepsilon^\mathrm{new}(\Delta t/2)$. 
This process is repeated for all other time grid points.

\section{Results and discussion}
\label{sec:results}
In the proposed scheme, the ultracold molecules are formed in one of the last 
bound levels, close to the dissociation limit. We have investigated
states from the whole range of the 
vibrational spectrum ($v=10$, $v=40$, $v=62$)
in order to gain a deeper understanding of the 
physical mechanism behind the control. Recall that the last bound level 
of the ground state of the sodium dimer has quantum number $v=65$.

Optimal pulses transferring all initial states to the target state ($v=0$)
were found if the guess pulse was chosen in a physically sensible way.
This means that the  pulse spectrum should coincide 
with the transition frequencies between vibrational levels in the ground
and excited state which have a good Franck-Condon overlap. The
transition frequencies with non-negligible Franck-Condon overlap range from
about 6000~cm$^{-1}$ to about 13000~cm$^{-1}$. This corresponds to a 
transform limited pulse of 5~fs full width half maximum (FWHM). 
However, not the whole spectral range of 7000~cm$^{-1}$ needs to be covered.
The Franck-Condon overlaps of the initial and the target state with the
vibrational levels of the excited state
indicate one or two spectral regions which should be important
in the transfer. It was sufficient to  accordingly choose one or
two central frequencies
of the guess pulse. Convergence to reach 99\% of the objective
could then be achieved with a relatively small number of iterations. 

\subsection{Simple search strategies}
\label{subsec:simple}
\begin{figure}[tb]
  \centering
  \includegraphics[width=0.99\linewidth]{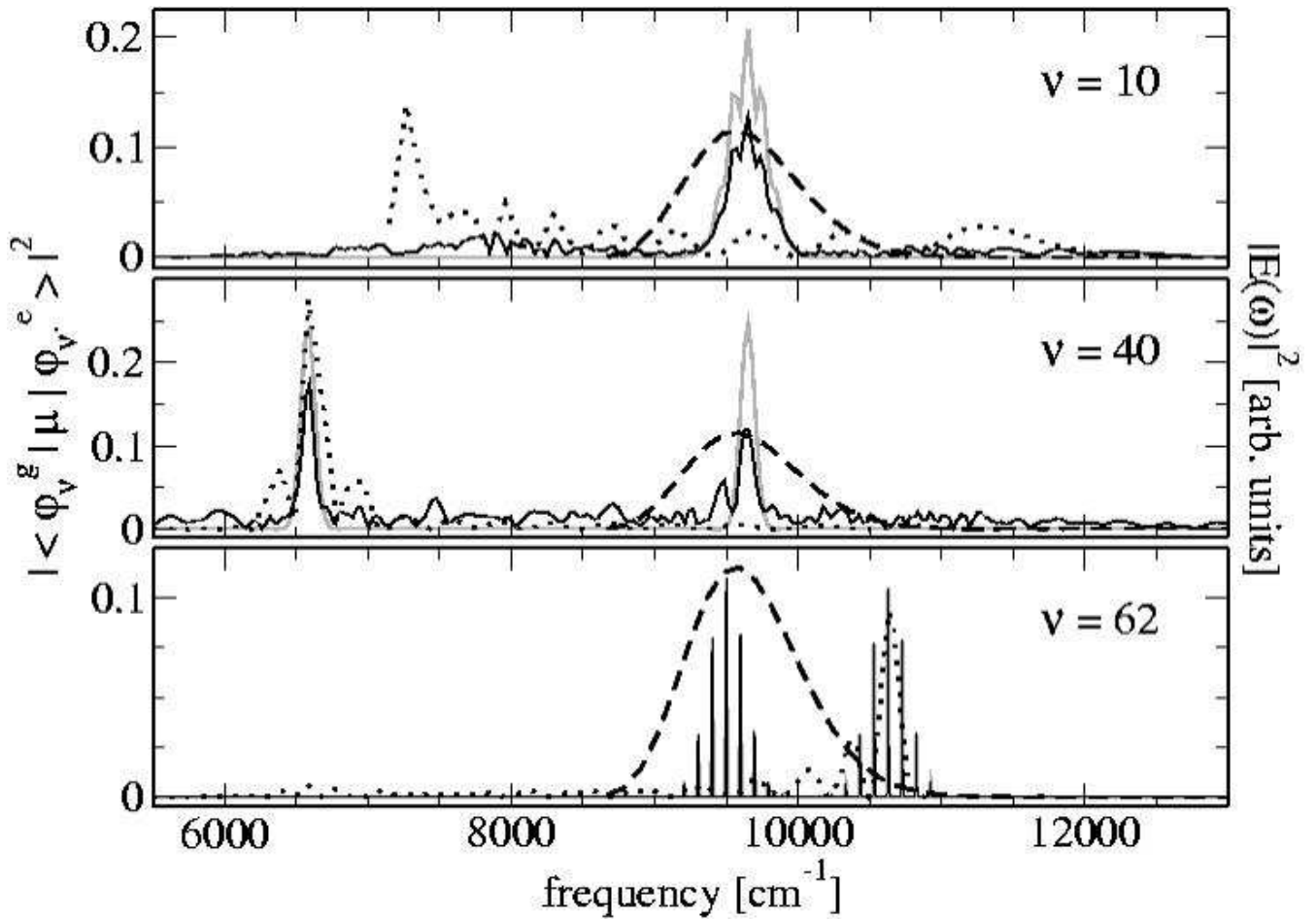}
  \caption{The Franck-Condon overlaps of the intial state ($v=10,40,62$, 
    dotted lines) and of the target
    state ($v=0$, dashed lines) with all excited
    state vibrational levels ($v^\prime$) are shown vs. frequency. 
    Also shown are the
    spectra of obtained optimal fields (solid black lines) and of the 
    corresponding guess fields (solid grey lines).}
  \label{fig:spec}
\end{figure}

In addition to the spectral range covered by
the pulse (cf. Fig.~\ref{fig:spec}), 
the rate of convergence depends on the two control knobs
of the optimal control algorithm, the intensity of the pulse or the
integrated pulse energy and the length $T$ of the optimization time 
interval. If we choose a comparatively high intensity and long time $T$
($T\ge T^* = 2 \pi /\omega = 2\pi h / |E_v - E_{v-1}|$
with $E_v$ the binding energy of level $v$), 
we allow the algorithm a lot of freedom. 
Convergence to 99\% of the objective is then reached very fast 
($N_\mathrm{it} \le 30$). 
This corresponds to a few hours of CPU time
for $v=10$ and $v=40$ and to a few days for $v=62$ on a
Linux PC (the increase for  $v=62$ is caused by the necessity
 to store the backward propagated
wave functions on disc rather than in memory for long time intervals).

\begin{figure}[tb]
  \centering
  \includegraphics[width=0.99\linewidth]{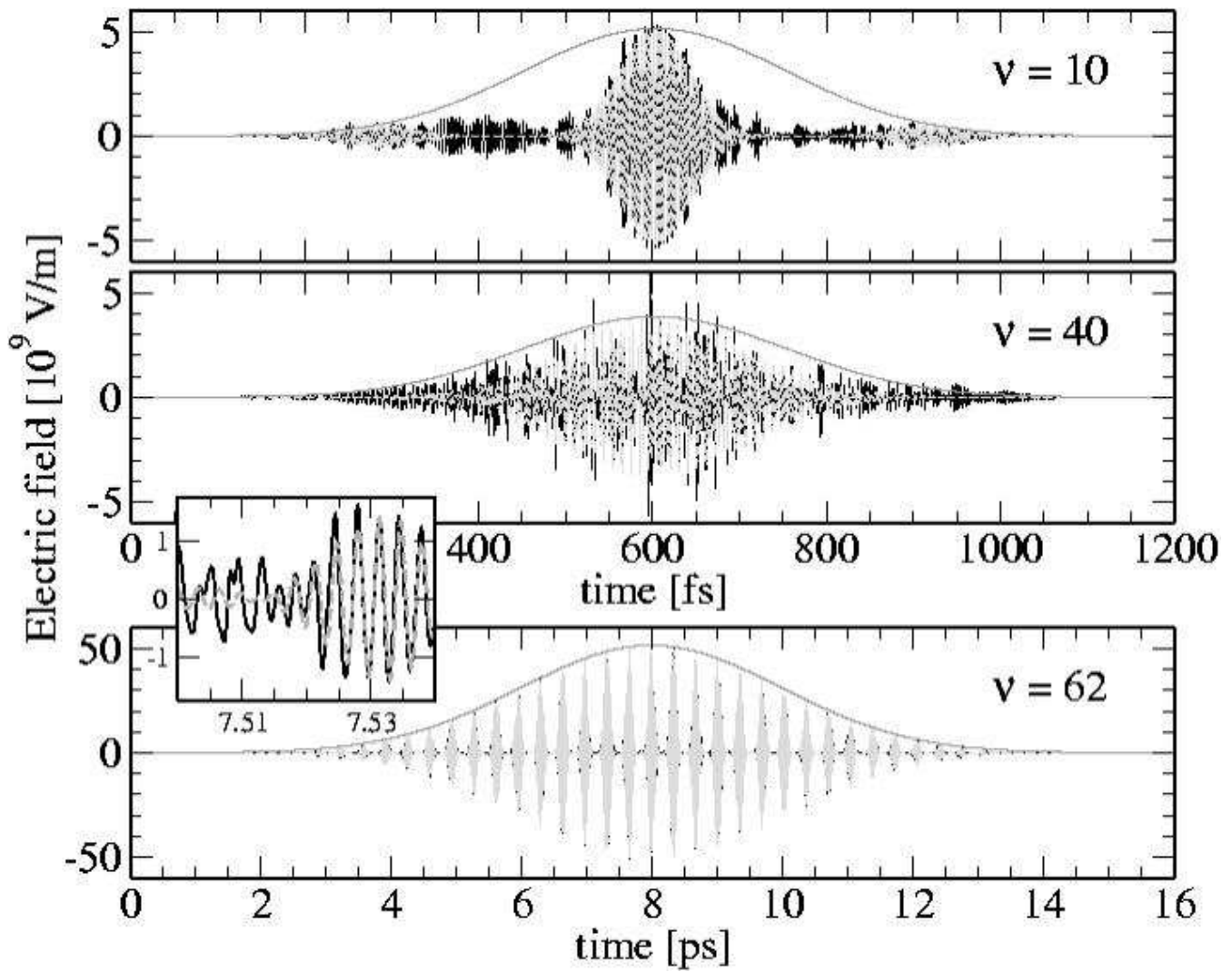}
  \caption{The optimal fields (black solid lines) 
    corresponding to Fig.~\ref{fig:spec}.
    Also plotted are the respective 
    guess fields (dashed lightgrey lines) and 
    the shape function $S(t)$ (solid darkgrey lines).
    The differences between optimal and guess fields 
    for $v=62$ can hardly 
    be seen on the time scale of 16~ps, therefore
    the inset shows the enlargement of a small interval.
  } 
  \label{fig:field}
\end{figure}
The resulting pulses are shown in Figs.~\ref{fig:spec} and \ref{fig:field}. 
A sequence of pulses with 100~fs FWHM was chosen for $v=10$ and $v=62$, while
a single pulse of 355~fs FHWM was taken as guess for $v=40$.
The spectra of the pulses (solid lines in Fig.~\ref{fig:spec})
are compared to the Franck-Condon overlaps
of the initial (dotted lines) and target (dashed lines) 
state with all vibrational levels of the
excited state potential. For the target state $v=0$,
the Franck-Condon overlaps with all excited state vibrational levels 
are characterized by a simple broad peak between 
8700~cm$^{-1}$ and 11000~cm$^{-1}$. The corresponding vibrational levels 
of the excited state have quantum numbers $v^\prime=1$ to $v^\prime=20$.

For initial state $v=10$, non-zero Franck-Condon overlaps can be found between
7150~cm$^{-1}$ and 12500~cm$^{-1}$ corresponding to levels 
$v^\prime=0$ to $v^\prime=54$. In this case, the control task
is comparatively simple: There is a number of excited state vibrational levels
which have good Franck-Condon overlap with both the initial and the target
state, i.e. a direct pathway exists. 
The simplest solution is therefore given by a pulse which addresses
the transition frequencies between  these "common" levels 
($v^\prime=1$ to $v^\prime=20$) and initial and target state. This 
solution is rapidly found by the optimal control algorithm:
The transition frequencies between  these "common" levels 
($v^\prime=1$ to $v^\prime=20$) and the target state are already contained in 
the guess pulse (grey line in Fig.~\ref{fig:spec}), and the
transition frequencies between the "common" levels and the initial state
around 7800~cm$^{-1}$ can clearly be seen in the spectrum of the optimal pulse. 
Since the guess pulse is rather intense, other pathways are excited as well.
If an initial guess which is even closer to the
direct pathway solution, i.e. a pulse with two central frequencies is employed, 
optimal solutions
can be found with less intensity and larger FHWM, i.e. 
smaller spectral bandwidth (not shown). Note, that even in this simple case
where  the solution basically can be guessed, the knowledge of only the
transition frequencies is not enough. Such a guess pulse 
with two central frequencies transfers less than
1\% from the initial to the target state while the time-frequency 
correlations of the optimal pulse allow for more than 99\% transfer.

The situation becomes more involved for higher excited initial states. No
direct pathways, i.e. no excited state vibrational levels which 
simultaneously have good
Franck-Condon overlap with initial and target state exist. Moreover,
non-zero Franck-Condon overlaps with the excited state vibrational levels
are usually found in different spectral regions for the initial and the target
state. The guess pulses are therefore chosen to contain two 
central frequencies addressing the relevant transition frequencies. 
If sufficient intensity is allowed, a solution can rapidly be found. 
The intensity is needed to find other pathways which are not contained
in the guess pulse, and the resulting spectrum is rather broad
(cf. Fig.~\ref{fig:spec}, middle panel). Note furthermore the different 
time scales in Fig.~\ref{fig:field} where the length of the time interval $T$
has been chosen to be at least twice the vibrational period of the initial
state. Since the level $v=62$ is very close to the dissociation 
limit, its binding energy is very small and its vibrational period is
very long. The required $T$ increases up to 240~ps for $v=65$. Such long
times obviously pose a problem,
this will be addressed in Sec.~\ref{subsec:T}.

\subsection{Restricting the pulse intensity}
\label{subsec:I}

Since optimal control is based on constructive and destructive interference
between different quantum pathways, the intensity of the pulse must be
sufficiently high to excite several such pathways. Furthermore the
guess pulse must contain sufficient intensity in the relevant modes for
the algorithm to find a solution. If the overlap of the 
initial state propagated with the guess field and the target state is very small in
 Eq.~(\ref{eq:deltaeps}), the changes in the field are also very small.
The algorithm then needs a very large number of iterations, or it may not
 converge at all.
Therefore there exists a minimum intensity necessary to find a solution.
Note that the minimum intensity we have found theoretically
does not necessarily have to be the same in a corresponding experiment. This is
due to the differents ways in which a solution is obtained theoretically and
experimentally. The relation between the two still needs to be clarified
which is, however, beyond the scope of the present work.

The integrated pulse energy is given by
\begin{equation}
  \label{eq:E_p}
  \mathcal{E}_P = \epsilon_0\, c\, A\,  
  \int_0^\infty |E(t)|^2 \Fkt{d}t 
\end{equation}
with $E(t)$ the optimal field, $A$ the area which is covered by the laser
($A=\pi r^2$ with $r=300\mu$m), 
$c$ the speed of light and $\epsilon_0$ the electric constant.
The integration can be done numerically over the optimal field or analytically
over the shape function, the latter usually being 
sufficient to obtain an estimate.
\begin{table}[htb]
  \centering 
  \begin{tabular}{|c|c|c|c|}
    \hline
    $v$ (initial state) &10 & 40 & 62\\ \hline
    $\mathcal{E}_\mathrm{pulse}$&  60~$\mu$J & 1.5~mJ & 3.9~mJ \\
    \hline
  \end{tabular}
  \caption{Minimum intensities needed to find a convergent solution transferring
  the initial state ($v=10,40,62$) to the ground state ($v=0$) }
  \label{tab:minI}
\end{table}
Table~\ref{tab:minI} lists the minimum pulse energy necessary
to find an optimal solution within $N_{it}\le 100$ iterations for
$v=10$, $v=42$, and $v=62$.
The increasing difficulty of the control task with increasing vibrational
excitation is reflected in the increase in required pulse energy. 

\begin{figure}[tb]
  \centering
  \includegraphics[width=0.99\linewidth]{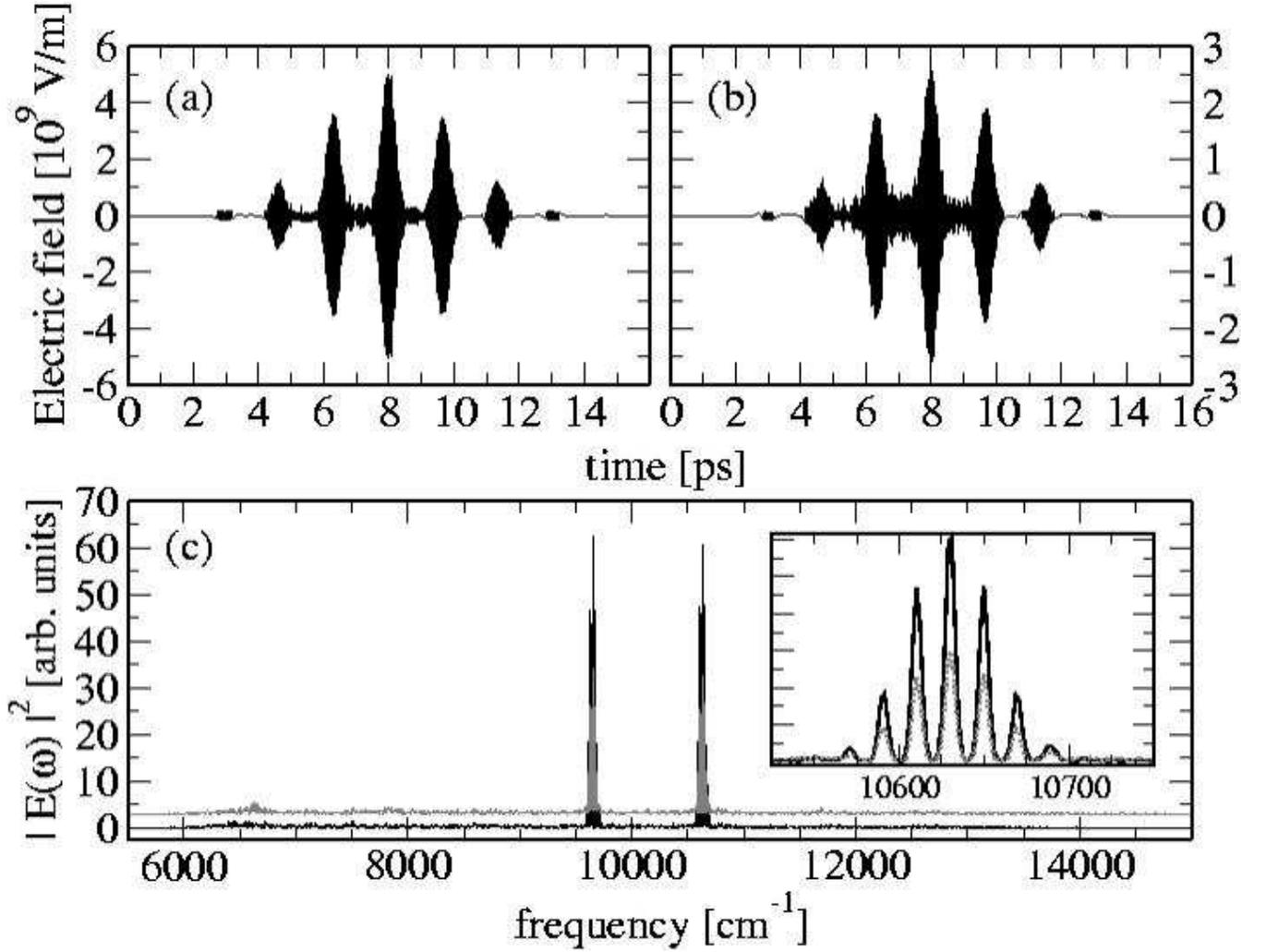}
  \caption{The fields (a-b) and spectra (c) of optimal fields for $v=62$.
    The spectrum of the stronger pulse (a) is plotted in black while the spectrum
    of weaker field (b), slightly shifted for visibility, is plotted in grey.
    The inset shows an enlargement of the high-frequency peak. 
  }
  \label{fig:field_minI}
\end{figure}

Since part of the intensity is necessary to find the relevant frequencies,
one can speculate that it should be possible to further reduce the intensity
by dividing an optimal field which contains already the relevant frequencies
by some factor and to restart the optimal 
control algorithm. This is indeed possible. New optimal solutions within
a reasonable number of iterations were found after dividing the 
optimal field of a previous calculation by a factor between two and five (reducing
the pulse energy by a factor between four and 25). Using an analytical guess
with the corresponding intensity did not lead to a convergent solution.
An example  is shown for $v=62$ in Fig.~\ref{fig:field_minI}.
The spectrum of the weaker field becomes broader, in particular 
new peaks appear at about 6600~cm$^{-1}$ and 7800~cm$^{-1}$.
The dynamics induced by the two optimal fields of Fig.~\ref{fig:field_minI}
has been analysed by calculating the time-dependent population of
vibrational levels, $|\langle \varphi_{v/v^\prime}|\varphi_{g/e}(t)\rangle|^2$. 
The number of levels which attain at some time $t$
a population of more than 10\% and 5\%, respectively, are given in 
Table~\ref{tab:pop_I}. 
\begin{table}[htb]
  \begin{tabular}{|c|c|c|}
    \hline
     &$E_0=0.01$~a.u. & $E_0=0.005$~a.u. \\
     &$\mathcal{E}_\mathrm{pulse}=16$~mJ & $\mathcal{E}_\mathrm{pulse}=4$~mJ \\ \hline
    $N^\mathrm{ground}_\mathrm{Pop>0.10}$ &3 &6 \\ \hline
    $N^\mathrm{ground}_\mathrm{Pop>0.05}$ &23 &13 \\ \hline
    $N^\mathrm{excited}_\mathrm{Pop>0.10}$ &7 & 7\\ \hline
    $N^\mathrm{excited}_\mathrm{Pop>0.05}$ &24 & 16\\ \hline
  \end{tabular}
  \caption{Number of vibrational levels of ground and
    excited state whose population exceeds 10\% 
    ($N_\mathrm{Pop>0.10}$) and 5\% ($N_\mathrm{Pop>0.05}$) 
    at some time $t$ ($0<t<T$). Note that the 7 excited state levels with more
    than 10\% population are not the same for 16~mJ and for 4~mJ.}
  \label{tab:pop_I}
\end{table}
The weaker pulse populates a smaller number of 
vibrational levels. It is interesting to note that the 
levels which get populated are not the same for the weaker and the stronger
pulse. Hence, a different solution is found after the reduction
of intensity. This is confirmed by the objective $F$ being
$F=7.9\cdot10^{-5}$ after the division of the optimal field
by a factor of two, but becoming
$F=0.97$ after 25 more iterations of the algorithm. 
The fact that the field is not simply scaled but a new solution needs to
be found, explains why the intensity of the field can only
be reduced by a factor between two and five, but not by one or two orders
of magnitude. Again, while some information is already contained in the field,
a certain minimum intensity is needed to excite other pathways which constitute
the new solution.

\subsection{Restricting the optimization time $T$}
\label{subsec:T}

The optimization time $T$ is a parameter of the calculations, and 
as explained in Sec.~\ref{subsec:simple} a good guess 
is given by taking $T$ to be twice the time corresponding to the smallest
frequency which occurs in the problem. Since we are interested in the
last bound levels, this time becomes very large (on the order of
nanoseconds for Na$_2$). 
Such a time scale is too long to efficiently compete with
collisions and spontaneous emission. 

Simply choosing a smaller value of $T$ does not solve the problem. In this case,
the optimal control algorithm did not find any solution for many different
guess pulses. A comparatively simple remedy to the problem was found
by using the result of an optimization with a large value of $T$,
restricting it in time, and employing this modified field 
as initial guess pulse for a second run of the optimization
algorithm. The restriction in time was done by Fourier-transforming the
field, deleting points in frequency space and Fourier-transforming it back to
the time domain. Keeping only every 10th value of the field in frequency space
reduces the time interval by a factor of 10. The shape function for the 
second optimization run was chosen to be $S(t)=\sin(\pi t/\tilde{T})$ 
where $\tilde{T}$ denotes the reduced time interval.

\begin{figure}[tb]
  \centering
  \includegraphics[width=0.99\linewidth]{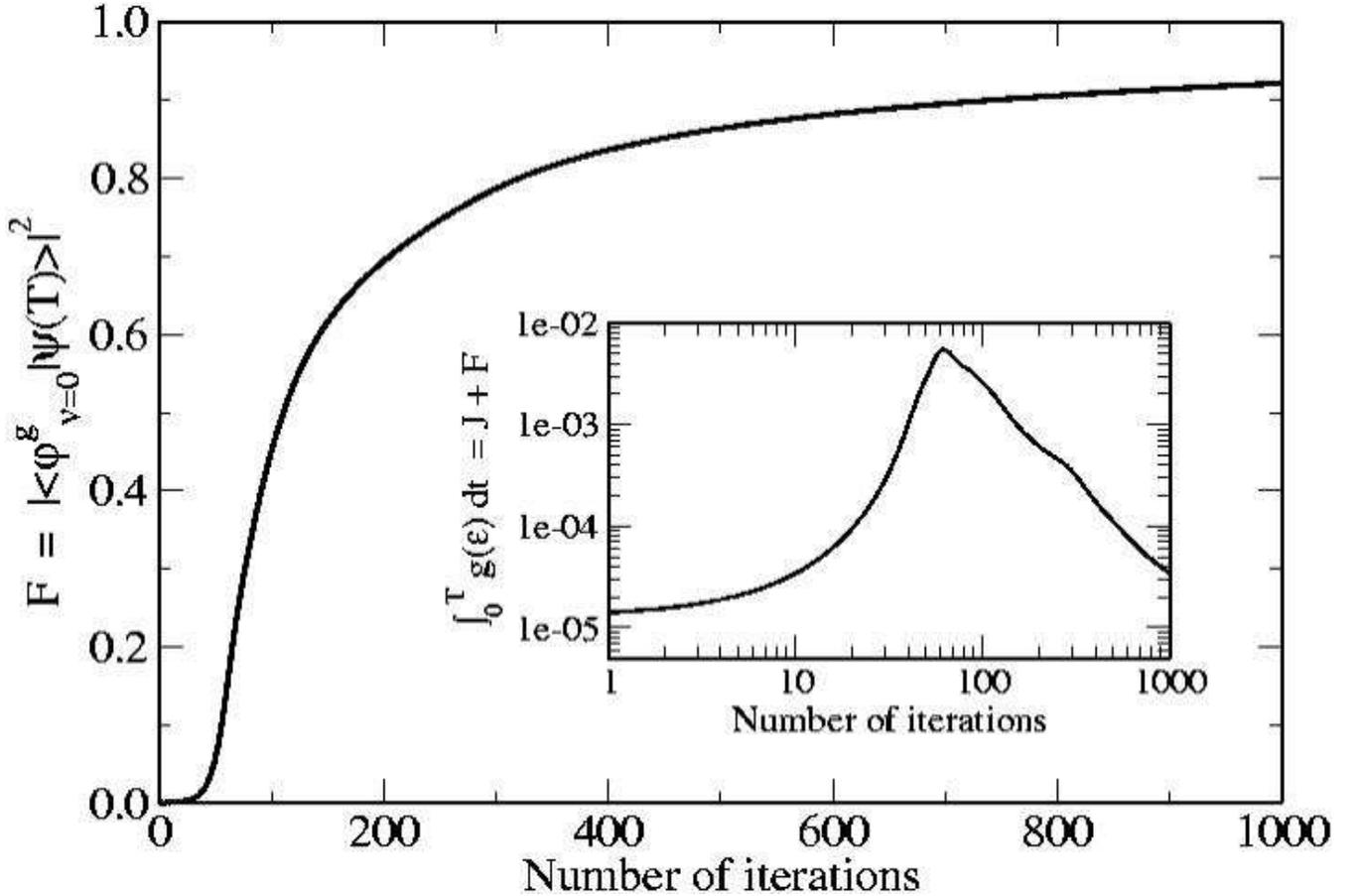}
  \caption{Convergence of the optimal control algorithm after the total time
    $T$ (corresponding to the vibrational period of the initial state)
    has been reduced by a factor of 8. 
  }
  \label{fig:conv}
\end{figure}

Such a reduction of the optimization time has been tested for up to a factor
of 20. Remarkably, it did not matter whether the points in frequency space
  were deleted symmetrically around $\omega=0$ or not. There is no rigorous
upper limit to the factor, by which $T$ can be reduced. However, the more $T$
is reduced, the slower becomes the convergence. There is furthermore no 
upper limit to the objective which can be reached with reduced $T$. This is
shown in Fig.~\ref{fig:conv} where the objective is plotted vs. the number
of iterations of the control algorithm. 
After reducing $T$ by a factor of 8,
a field which transfers more than 90\% of the initial state to $v=0$ is
found after  $N_\mathrm{it} \sim 1000$ iterations.
Note, that each iteration requires a full time propagtion. 
Nevertheless, the calculations do not become prohibitively expensive due to
the shorter propagation time ($T/8$).
The inset of Fig.~\ref{fig:conv} shows $\int_0^T g(\varepsilon) \Fkt{d}t$ 
vs. the number of iterations.
The integral corresponds to the value by
which the objective is increased at each iteration, 
While this value decreases algebraically after about 60 iterations, it is still
sufficient to increase the objective from about 20\% after $N_\mathrm{it}=60$ to
more than 90\% after $N_\mathrm{it}=1000$. The larger number of required
iterations is not only due to the restriction in time, but also due to a
reduction in integrated pulse energy and hence intensity.

\begin{figure}[tb]
  \centering
  \includegraphics[width=0.99\linewidth]{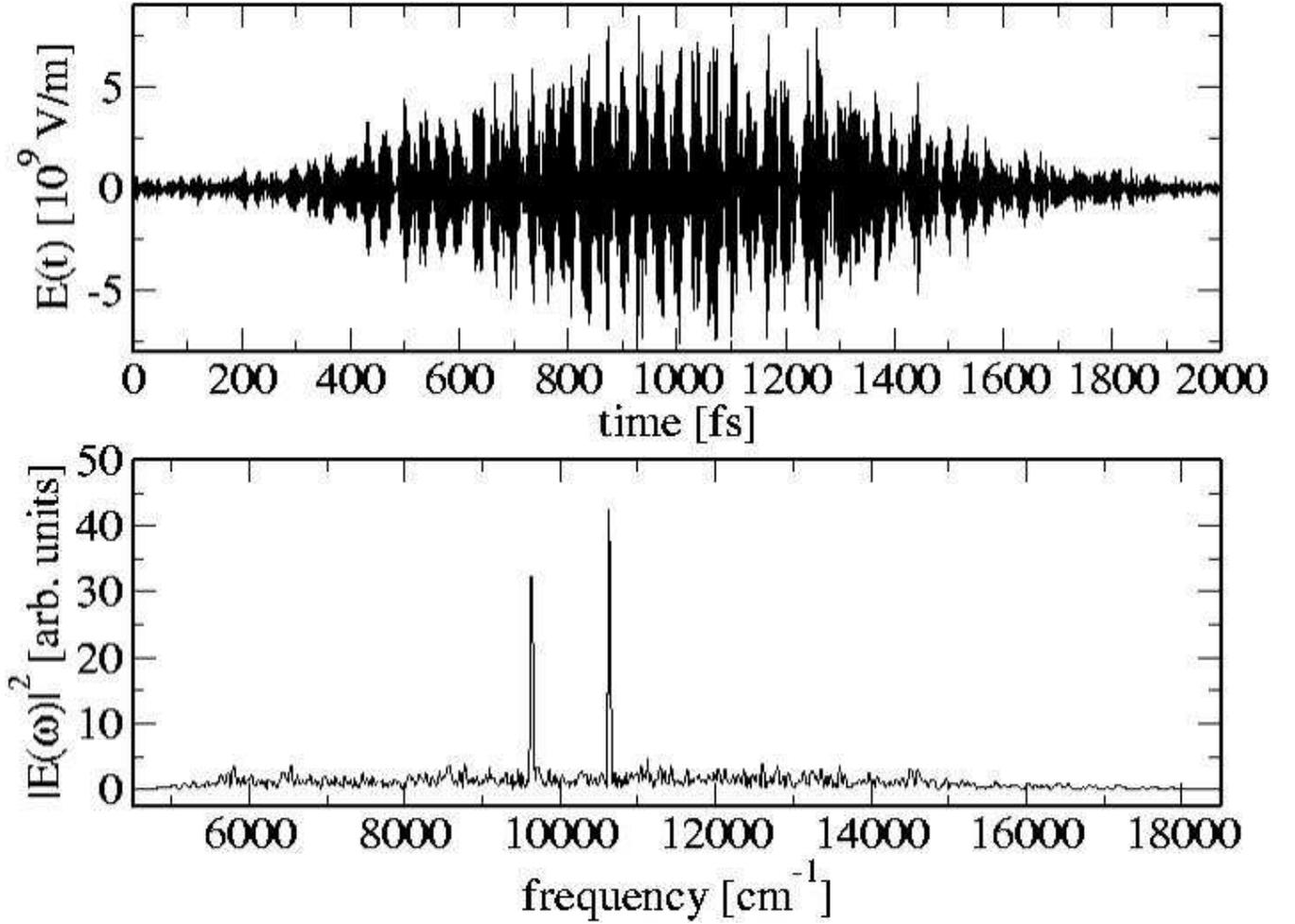}
  \caption{The optimal field and its spectrum. The optimization time 
    interval has been reduced by
    a factor of 8.
  }
  \label{fig:field_T}
\end{figure}

An example of an optimal field and its spectrum after $N_\mathrm{it}=1000$
are shown in Fig.~\ref{fig:field_T}.
The field shows a sequence of very short sub-pulses which is characteristic
for optimal fields~\cite{HornungPRA02a}. The spectrum is rather broad, 
covering about 9000~cm$^{-1}$. It should be noted, however, that already
the spectrum after the first optimization run with large $T$ was rather 
broad. At this point it is very difficult to decide whether this broadness
is physically necessary or or whether it is
an arteficial by-product of the mathematical algorithm.
Unfortunately, it is not possible in a straightforward way 
 to include a restriction in
bandwidth as condition in the algorithm without ruining the 
property of monotonous
convergence (see App.~\ref{sec:restrict} for a more detailed discussion).

\begin{figure}[tb]
  \centering
  \includegraphics[width=0.99\linewidth]{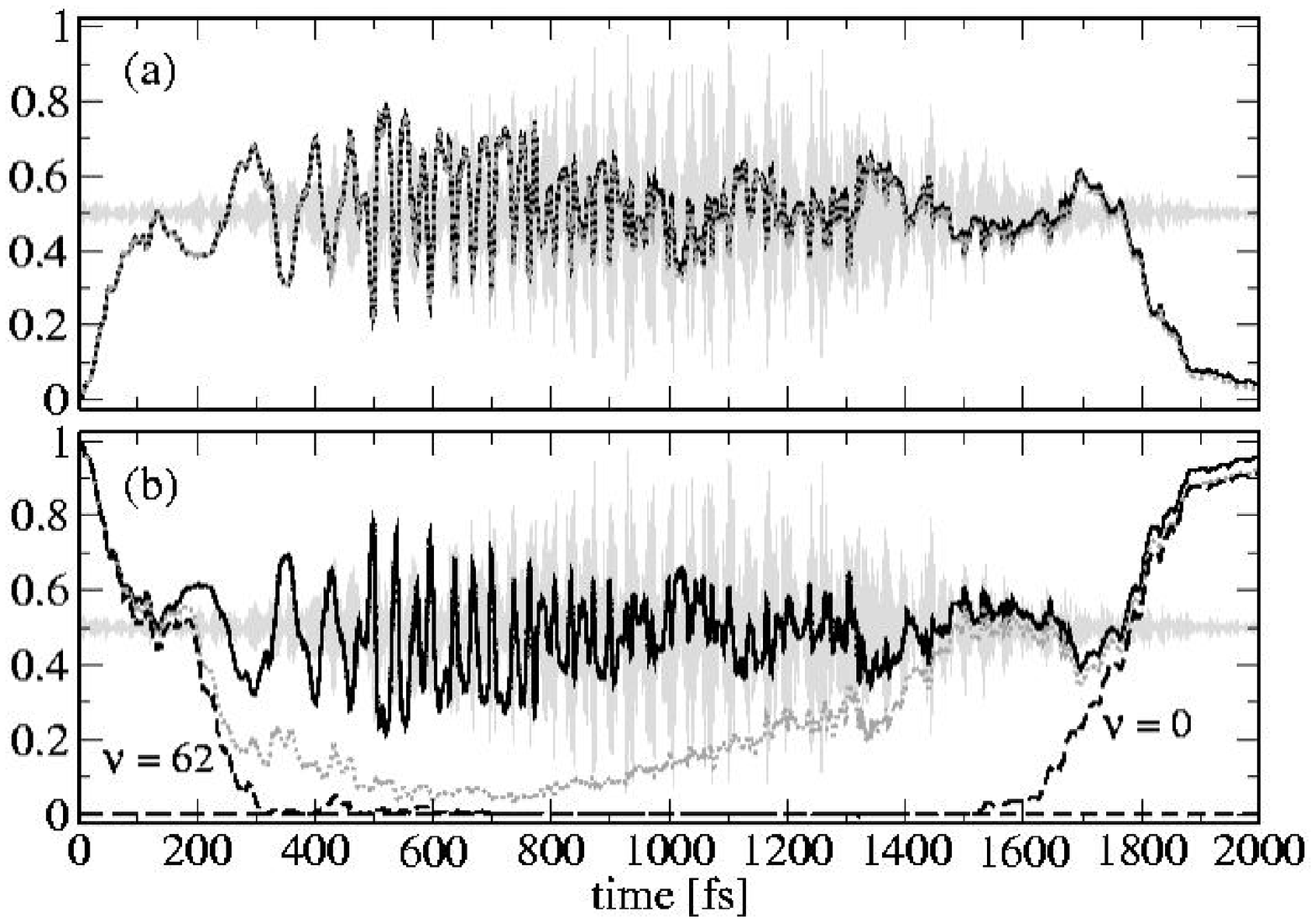}
  \caption{Population analysis of the dynamics with the optimal field for the
    electronically excited state (a) and the ground state (b). The total
    population $\big| \langle g/e | \Psi(t)\rangle\big|^2$
    (solid black line) is compared to the sum of populations
    of all {\em bound} vibrational levels 
    $\sum_v |\langle \varphi^{g/e}_v | \Psi(t)\rangle|^2$ 
    (dotted grey line) with the field plotted in the background (lightgrey, 
    scaled for comparison). For the excited state, the black and the grey line
    are almost identical indicating that the continuum is not  populated,
    while the curves differ for the ground state indicating 
    significant population of the continuum at {\em intermediate} times.
    Panel (b) shows furthermore the population of
    initial ($v=62$) and target ($v=0$) vibrational level (dashed lines).
  }
  \label{fig:pop_T8}
\end{figure}

\begin{figure}[tb]
  \centering
  \includegraphics[width=0.99\linewidth]{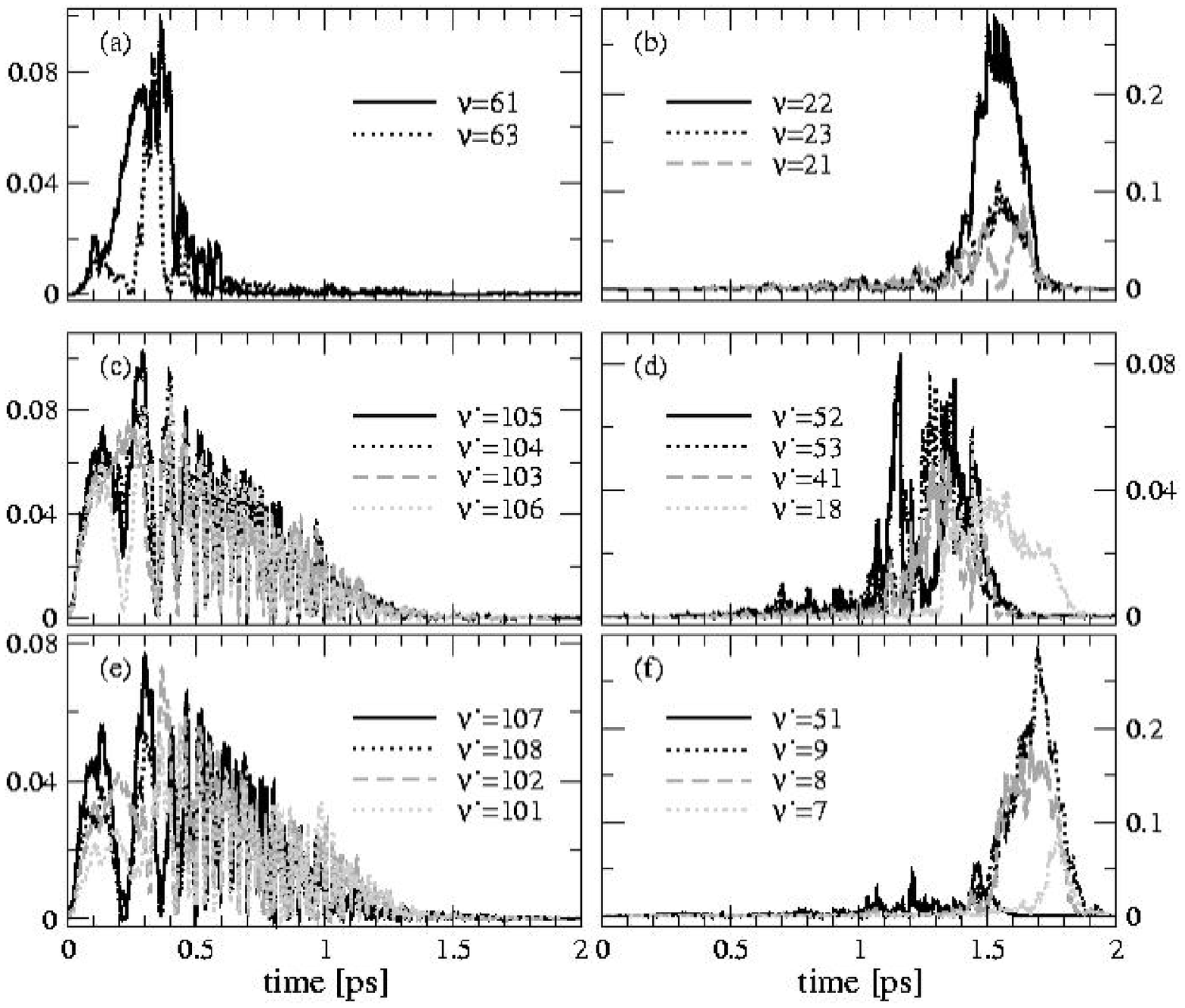}
  \caption{Vibrational level population of the dynamics with 
    the optimal field for the
    electronically excited state (c)-(f) and the ground state (a)-(b)}
  \label{fig:vib_T8}
\end{figure}

More insight is gained by analyzing which vibrational levels 
 the optimal field populates during the course of time.
Fig.~\ref{fig:pop_T8} shows the total population of the two channels
(solid black lines) with the optimal field (in grey) 
which has been scaled for comparison.
The sub-pulse structure of the field 
corresponds to population cycling between ground and
electronically excited state. For the ground state, the population of
initial ($v=62$) and target ($v=0$) vibrational level (dashed lines)
are also plotted. Finally, the dashed grey lines in Fig.~\ref{fig:pop_T8}
show the sum of populations of all vibrational levels, i.e.
$\sum_v |\langle \varphi^{g/e}_v | \Psi(t)\rangle|^2$. For the 
electronically excited state, the solid black and dashed grey lines overlap,
i.e. only bound levels get excited. This is not surprising: The highest
vibrational levels which get populated (Cf. Fig.~\ref{fig:vib_T8})
are those which have good Franck-Condon overlap with the initial state. 
These levels are still comparatively strongly bound ($v^\prime=101\ldots108$
out of about 220 bound levels). For the ground state, the solid black and
dashed grey lines differ remarkably, corresponding to a non-negligible
population of the continuum. However, this does not need to worry us.
The continuum gets populated at intermediate times for very a short period
when the Na$_2$ system is in a {\em coherent} superposition with
the optimal field. The corresponding population is therefore not lost but
brought back to bound levels at later times and eventually transferred to
$v=0$. Fig.~\ref{fig:vib_T8} shows the population of those vibrational
levels vs. time which acquire more than 5\% of the overall population
at some time $0 \le t\le T$. Ground state vibrational levels are
displayed in the upper panel, while excited state vibrational levels are
shown in the middle and lower panel.
The overall number of vibrational levels which get populated is
7 for the ground state ($v=62$ and $v=0$ have been plotted in 
Fig.~\ref{fig:pop_T8} for better visibility) and 16 for the electronically
excited state. For the ground state (Fig.~\ref{fig:vib_T8} (a)-(b)), 
the two neighbouring levels of the initial
state get populated at early times. At intermediate times, the population
is spread over many different levels (and the continuum) while toward the
end of the time interval, the population is concentrated in three levels
($v=21,22,23$). None of this levels is populated for more than 300~fs.
This is the time which needs to be compared to the time scale of 
vibrational energy pooling. For the excited state
(Fig.~\ref{fig:vib_T8} (c)-(f)), the vibrational levels which get 
populated can be assigned to three different groups: At early times 
a number of levels ($v^\prime=101\ldots 108$) with good Franck-Condon
overlap with the initial state get populated. While the population 
cycles back and forth between ground and excited state, these levels
are overall populated for a comparatively long time. 
After about 1~ps, intermediate
levels ($v^\prime=18,41,51\ldots 53$) are populated. Toward the end of the time
interval the population is concentrated in levels ($v^\prime=7,8,9$) which have
a very good Franck-Condon overlap with the target state, $v=0$. Also none
of the excited state vibrational levels is populated for more than
300-400~fs which should be short enough to compete with vibrational
energy pooling.

Note, that which specific vibrational levels are populated differs for
different optimal pulses which are the result of initial guess pulses
with different frequencies and intensities and of reduction of $T$
by different factors. The overall scheme of population, in particular
for the excited state with the three groups at early, intermediate and long 
times, is found in all cases. This is not surprising because it simply
reflects the fact that a multi-step scheme is needed to transfer
a very weakly bound state to the vibrational ground state. 

\section{Conclusions}
\label{sec:concl}

We have shown that within a two-state model of Na$_2$ optimal fields
which transfer highly excited vibrational states to the vibrational 
ground state to more than 99\% can be found. We have performed
calculations for moderately, highly and and extremely highly excited
initial states ($v=10,40,62$ out of 66 bound levels). For moderately
excited vibrational levels, a direct pathway from the initial to the
target state exists, and a single central frequency in the guess field
was sufficient to obtain an optimal field. For higher excited levels,
a multistep scheme is required since initial and target state have
good Franck-Condon overlap with different vibrational levels of the
excited state. This was accounted for by assuming two central 
frequencies in the guess field. Experimentally, pulses with two 
central frequencies can be realized by sending a pulse with a single 
central frequency through a parametric amplifier.

The two control knobs of the optimization algorithm are the pulse
intensity and the optimization time. A comparatively high
intensity is needed to excite several quantum pathways whose 
interference constitutes the solution. The optimization time
is usually chosen to correspond to the longest vibrational period
of the system. We have explored ways to restrict both intensity
and optimization time. This was possible by using information
from optimal fields of a previous run of the algorithm
with large intensity / long optimization time in 
the guess field which was used as input in a second run of
the algorithm with smaller intensity and shorter optimization time. 
Optimal fields with integrated pulse energy on the
order of milliJoule and overall duration on the order of a few
picoseconds have thus been obtained.

The optimal control task we have investigated in this paper
is much more difficult than those of previous control studies 
on the sodium dimer, such as
populating certain ionization channels~\cite{PaciJCP98}
or displacing the ground state Gaussian wave 
packet~\cite{ShenCPL03}. The fact that solutions could still  be found, is
 due to the modification
of the optimal control algorithm to calculate the change in field rather
than the new field itself~\cite{JosePRA03}.
This ensures a smooth convergence of the field toward the optimal one
and renders the algorithm very powerful.

We have neglected several possible loss channels. Two-photon transitions
could populate an autoionizing state. However, since molecular 
autoionization occurs at small distances, 
the cross-sections are very small~\cite{HuynhPRA98}. 
The field could moreover transfer
angular momentum to the system thus exciting rotations. Both loss channels
can be modelled by adding one or a few electronic states to the model.
In general, additional channels do not necessarily pose a problem, 
they may even lead to more
participating pathways improving the control~\cite{PaciJCP98}.
Intuitively, the field is able to avoid population of these loss channels if 
its phase is orthogonal to the transition dipole. It has been shown
in local optimization calculations~\cite{AllonJCP93,MalinovskyCPL97}, 
that this condition on the phase of the
field can be used to lock the population in the desired 
channels.
In global optimization, convergence can be facilitated by imposing a penalty
on the loss channels~\cite{SklarzPriv}. 
Multiphoton ionization of the sodium atoms is 
more dangerous than excitation of rotations and molecular autoionization. It
is rather likely to occur at the intensities we have found. Our calculations
should therefore be repeated 
by adding an $N$-level Hamiltonian  describing the atomic levels
to the model. The task of the modified control problem is to
transfer the vibrationally excited molecules to $v=0$ while
avoiding population of the ionization 
loss channels at {\em any} time during the optimization 
interval. The latter has to be formulated as an additional constraint. 
The function $g(\varepsilon)$ does then also depend on the state $\varphi$, 
$g(\varepsilon,\varphi)$, and the equation for the new field 
has to be rederived. Work in this direction is in progress.

In a recent experiment~\cite{XuPRL03}, ultracold Na$_2$ molecules have been 
produced from an atomic sodium condensate with the molecules occupying 
the $v=14$ vibrational level of the  a$^3\Sigma_u^+$ state. This vibrational
level is very weakly bound ($<1$ cm$^{-1}$), and at such small binding energies 
the hyperfine interaction mixes singlet and triplet states~\cite{SamuelisPRA00}.
Note that in the asymptotic region
the wave function of the triplet $v=14$ level is identical
to that of the singlet $v=64$ level.
The coupling between hyperfine states 
implies that also vibrational levels of the singlet state
are populated, hence our results considering the X$^1\Sigma_g^+$ should be applicable
to the MIT experiment. Second, to obtain quantitative results
our model should be improved to include all the states which are coupled
by hyperfine interaction. This requires, however, several technical improvements
and is therefore the subject of future work.

The MIT experiment raises another question regarding the
feasibility of optimal control experiments with ultracold molecules.
The reported lifetime of the molecular cloud is a few ms~\cite{XuPRL03}.
Conventional pulse shapers operate at about 1 kHz resulting in one
cycle of the learning loop per millisecond. In standard control experiments
 a few thousand cycles  can therefore be performed in a comparatively 
short time. If the MIT experiment were to be combined with a control
scheme, only one or two cycles could be performed before the molecules
are lost from the trap. It therefore seems necessary to provide more
information from theoretical considerations than simply 
the required spectral bandwidth and intensity of the pulse. 
A promising approach has been suggested by de Vivie-Riedle and 
coworkers~\cite{HornungJCP01} who defined the mask pattern of the
optimal pulse by discrete Fourier transforms. 
Provided the calculated pulse is not too complex, 
this mask pattern can be directly fed into the pulse shaper
avoiding the many cycles of a learning loop.

In addition to accounting for ionization losses and hyperfine interaction,
our two-state model should be improved to consider rotations. This is 
particularly important because the molecules are not necessarily created
with $J=0$ in the experiments using Feshbach resonances or three-body 
recombination. Since the rotational excitation is rather small, it will be
sufficient to treat it by additional electronic states, for which the
centrifugal barrier for each $J$ is added to the atom-atom interaction.
One may speculate that some rotational excitation may even facilitate 
the stabilization process since the initial wave function 
due to the centrifugal barrier is more localized than the one for $J=0$.

Another possibility which leads to a more confined initial state
for the stabilization process is given by forming the molecules
by photoassociation.  Two-color photoassociation or
spontaneous emission after one-color photoassociation
 can directly populate levels around $v=40$. 
We have shown in our calculations that the optimal control 
task for such vibrational states
which are still highly excited but already well-bound is much easier
than for the last bound levels. The required intensities are hence
more moderate (Cf. an integrated pulse energy of 1.5~mJ for $v=40$ vs. 
4~mJ for $v=62$). 

Finally, our calculations prompt the question whether
spectrally simpler fields than those obtained in this study
can be guessed. Our population analysis
has shown that at least two cycles through the excited state are
required. The first cycle populates excited state vibrational levels
which have good Franck-Condon overlap with the initial state, and the
second populates excited state vibrational levels with good Franck-Condon
overlap with the target state. One could hence divide these steps,
optimizing each separately. First calculations show that 
the intensities required for one such step are much lower than the ones
completing the control task at once. Note that while the intensity
is reduced, the overall time (including all steps) is prolonged. 
Another possibility is given
by finding a solution for each of the steps intuitively without
the optimal control algorithm, for example chirping the sequence
of pulses. Furthermore, stabilization via optimal control could be 
combined with chirped pulse photoassociation~\cite{JiriPRA01,Luc04}.
For the Cesium dimer, it has been shown that a spatially focussed wave packet
can be formed on the excited state~\cite{Luc04}. To stabilize such a
state to the ground vibrational level of the electronic ground state
via OCT should also require comparatively little intensity.
Our results furthermore indicate that STIRAP or any other adiabtic process
is not likely to be a 
successful tool in transferring the highly excited molecules to the
vibrational ground state. The multistep scheme implies that there is no
direct adiabatic path, i.e. the Franck-Condon overlaps of a direct
path are extremely small. Correspondingly, 
the intensity required for a direct adiabatic path is expected to be orders of 
magnitude larger than that obtained by us for an optimal field.
Adiabaticity moreover requires the transfer to be slow. If we
quantify slow as being at least one order of magnitude slower than the
longest vibrational period of the problem, the process has to occur
on the time scale of nanoseconds in the case of Na$_2$ and 
even longer times for heavier molecules. These are the time scales
where loss processes become important and where condensate dynamics and
molecular dynamics become entangled.

In conclusion, we have applied in this study optimal control
theory to the stabilization of ultracold molecules. While we have started
with a comparatively simple model, our present results point toward
several directions of improvements which in turn correspond to different
experimental schemes. We thus hope that the present work stimulates
further research uniting the fields of cold molecules and optimal 
control.

\begin{acknowledgments}
  We would like to thank the Cold Atoms and Molecules team of 
  Laboratoire Aim\'e Cotton, in particular N. Vanhaecke, O. Dulieu and P. Pillet,
  for fruitful discussions. C.P.K. is grateful to T. Kl\"uner for 
  assistance and to the Deutsche Forschungsgemeinschaft for
  financial support.
  This work has been supported by the European Union in the frame of the 
  Cold Molecule research training network under contract HPRN-CT-2002-00290
  and by the Israeli Science Foundation. 
  The Fritz Haber Center is supported
  by the Minerva Gesellschaft f\"{u}r die Forschung GmbH M\"{u}nchen, Germany.
\end{acknowledgments}

\appendix

\section{Brief review of Optimal control theory with the Krotov method}
\label{sec:krotov}

The Krotov method is one of several methods to derive an iterative algorithm
which connects the objective of optimal control, calculated at the final time
$t=T$, with the knowledge of the state of the system at intermediate times 
$0<t<T$. A general review was given in Ref.~\cite{SklarzPRA02}, while
a review for linear problems in the density matrix formalism was reported
in Ref.~\cite{JosePRA03}. In this paper, we need an even simpler version, that is
the Krotov method for linear optimization of state-to-state transitions which
we sketch in the following.

The equation of motion of the system is the Schr\"odinger equation,
\begin{displaymath}
  \frac{\Fkt{d}}{\Fkt{d}t} |\varphi(t)\rangle = -\frac{i}{\hbar} \Op{H}(\varepsilon)
  |\varphi(t)\rangle \,,
\end{displaymath}
whose right-hand-side is abbreviated by
\begin{equation}
  \label{eq:f}
  |f(t,\varphi(t),\varepsilon) \rangle = -\frac{i}{\hbar} \Op{H}(\varepsilon)
  |\varphi(t)\rangle\, .
\end{equation}
We omit the ket notation when just indicating the dependence on 
$|\varphi(t)\rangle$.
The starting point is now the modified objective $J$ of Eq.~(\ref{eq:J}), which
is a functional of the state of the system
and the field, $J[\varphi,\varepsilon]$. Due to integral term in Eq.~(\ref{eq:J}), 
$J$ is a functional of
the system's state at {\em all} times, 
i.e. a functional of the system "trajectory". The problem is now to find 
conditions on $J$ which maximize the objective $F$.

To this end, one introduces an auxiliary functional $L[\varphi,\varepsilon;\Phi]$,
\begin{equation}
  \label{eq:L}
  L[\varphi,\varepsilon;\Phi] = G(\varphi(T))-\Phi(0,\varphi_i) - 
  \int_0^T R(t,\varphi(t),\varepsilon) \Fkt{d}t
\end{equation}
with 
\begin{equation}
  \label{eq:G}
  G(\varphi(T)) = -F(\varphi(T)) + \Phi(T,\varphi(T)) \,,
\end{equation}
\begin{equation}
  \label{eq:R}
  R(t,\varphi(t),\varepsilon) = -g(\varepsilon)+\frac{\partial \Phi}{\partial t} +
  \mathfrak{Re}
  \left\langle\frac{\partial \Phi(t,\varphi(t))}{\partial \varphi} 
  \bigg| f(t,\varphi(t),\varepsilon)\right\rangle
\end{equation}
and  $\Phi=\Phi(t,\varphi)$ an arbitrary continuously differentiable real
function~\cite{SklarzPRA02,JosePRA03}. Recall that 
$|\varphi_i\rangle=|\varphi(t=0)\rangle$,
and $|\varphi_f\rangle$ is the target state.
Note that the derivative of $\Phi$ with respect to the state  means 
derivative with respect to both real and imaginary part of $|\varphi\rangle$,
$|\frac{\partial}{\partial \varphi}\rangle = 
| \frac{\partial}{\partial \varphi_R} + i \frac{\partial}{\partial \varphi_I}\rangle$.
If the state $|\varphi(t)\rangle$ and the field $\varepsilon(t)$
obey the Schr\"odinger equation, $R$ is given by 
$R=-g(\varepsilon)+\frac{\Fkt{d} \Phi}{\Fkt{d} t}$. It can then be 
shown~\cite{SklarzPRA02} that $L[\varphi,\varepsilon;\Phi]=J[\varphi,\varepsilon]$
for any scalar function $\Phi$, i.e. the minimization of $J$ is equivalent to the
minimization of $L$. In the latter one has complete freedom in the choice of 
$\Phi$. This property is used to construct an iterative algorithm. One first chooses
$\Phi$ such that $L$ is maximum with respect to $|\varphi(t)\rangle$, 
i.e. the worst case.
A new field is then derived from the condition of maximizing $R$ which in turn
leads to minimization of $L$. 

The first step (maximization of $L$) is equivalent to maximizing $G$ while 
minimizing $R$. This step is taken  at the old state, $|\varphi(t)\rangle^n$.
To simplify the equations,  the maximum and mininum 
conditions are  relaxed to extremum conditions. An additional 
condition in the end has then to ensure
that the algorithm indeed improves the objective in each 
iteration. The extremum conditions are written as
\begin{eqnarray}
  \left |\frac{\partial R(t,\varphi(t),\varepsilon)}{\partial \varphi} 
  \right\rangle_{|\varphi\rangle^n}
  &=& 0\,, 
  \label{eq:extremR} \\
  \left |\frac{\partial G(\varphi(T))}{\partial \varphi(T)} 
  \right\rangle_{|\varphi\rangle^n}
  &=& 0\,. \label{eq:extremG}
\end{eqnarray}
The partial derivatives are again with respect to both real and imaginary part
of $|\varphi\rangle$.
Eq.~(\ref{eq:extremR}) leads to
\begin{eqnarray*}
  0 &=& \left|\frac{\partial }{\partial \varphi}\right\rangle \left[ -g(\varepsilon) + 
    \frac{\partial \Phi(t,\varphi(t)) }{\partial t}  
    +  \mathfrak{Re}\left\langle\frac{\partial \Phi(t,\varphi(t))}{\partial \varphi} 
    \bigg| f(t,\varphi,\varepsilon)\right\rangle
  \right]_{|\varphi\rangle^n} \\
  &=&  \frac{\partial }{\partial t} 
  \left|\frac{\partial \Phi}{\partial  \varphi}\right\rangle_{|\varphi\rangle^n} +  
  \frac{1}{2}  \left|\frac{\partial }{\partial  \varphi} \right\rangle
  \left\langle\left(\frac{\partial\Phi}{\partial \varphi}\right)_{|\varphi\rangle^n} 
    \bigg|f(t,\varphi,\varepsilon)\right\rangle + \frac{1}{2} 
  \left\langle\left(\frac{\partial\Phi}{\partial \varphi}
    \right)_{|\varphi\rangle^n} \bigg|
     f(t,\varphi,\varepsilon)\right\rangle
       \bigg\langle \frac{\partial  }{\partial \varphi}  \bigg| \\
       && \;\;\;\; \;\;\;\; \;\;\;\; \;\;\;\; \;\;\;\; 
       + \frac{1}{2}  \left|\frac{\partial }{\partial  \varphi} \right\rangle
       \left\langle f(t,\varphi,\varepsilon)\bigg|
       \left(\frac{\partial\Phi}{\partial \varphi}\right)_{|\varphi\rangle^n} 
     \right\rangle   +        \frac{1}{2} 
       \left\langle  f(t,\varphi,\varepsilon) \bigg|
         \left(\frac{\partial\Phi}{\partial \varphi}  \right)_{|\varphi\rangle^n}        
       \right\rangle
       \bigg\langle \frac{\partial  }{\partial \varphi}  \bigg|
     \,.
\end{eqnarray*}
Using that $|f\rangle=\Fkt{d}/\Fkt{d}t|\varphi\rangle$ and calculating 
$|\partial /\partial \varphi\rangle\langle f| $ from Eq.~(\ref{eq:f}), we obtain
\begin{equation}
  \label{eq:EoMaux} 
  \frac{\Fkt{d}}{\Fkt{d} t}
  \left|\frac{\partial\Phi}{\partial \varphi}(t)\right\rangle_{|\varphi\rangle^n}  =
  -\frac{i}{\hbar} \Op{H}^+(\varepsilon) 
   \left| \frac{\partial \Phi}{\partial \varphi}(t)\right\rangle_{|\varphi\rangle^n}\,.
\end{equation}
Eq.~(\ref{eq:extremG})  together with the definition of $G(\varphi_f)$
(Eq.~\ref{eq:G})) gives
\begin{equation}
  \label{eq:initaux}
  \left|\frac{\partial \Phi}{\partial \varphi}(T)\right\rangle_{|\varphi\rangle^n}  = 
  -\left|\frac{\partial F(\varphi)}{\partial \varphi} (T)\right\rangle \,.
  \end{equation}
Recall that $F$ is the original objective, and $|\partial F/\partial \varphi\rangle$
can thus be explicitly calculated.
Eq.~(\ref{eq:EoMaux}) can then be interpreted as first 
order differential equation for 
$|\gamma(t)\rangle = 
\left|\frac{\partial\Phi}{\partial\varphi}(t)\right\rangle_{|\varphi\rangle^n} $ 
with the "initial" condition (at time $T$) given by
Eq.~(\ref{eq:initaux}). The knowledge of Eqs.~(\ref{eq:EoMaux}) and (\ref{eq:initaux})
is therefore sufficient to determine $\Phi(t,\varphi)$ to first order in 
$|\varphi(t)\rangle$,
\begin{equation}
  \label{eq:Phi}
  \Phi^{(1)}(t,\varphi) = 
  \mathfrak{Re}\left\langle\left(\frac{\partial \Phi}{\partial \varphi}(t)
    \right)_{|\varphi\rangle^n} \bigg|
  \varphi(t)\right\rangle =
  \frac{1}{2}\left[\langle\gamma(t)\; |\varphi(t)\rangle + 
  \langle\varphi(t)|\gamma(t)\rangle\right]\, \,.
\end{equation}
From $\Phi^{(1)}$, $G$ and $R$ can be constructed to first order, $G^{(1)}$ and 
$R^{(1)}$. Thereby the first 
step of the algorithm for a {\em linear} optimization problem is completed.
Note that $\Phi^{(1)}$ is never explicitly calculated, 
the equation of motion for $|\gamma(t)\rangle$ (Eq.~(\ref{eq:EoMaux})) will
be solved instead.

To realize the second step, $R^{(1)}$ must be maximized with respect to the field.
This step is supposed to lead to a new field, $\varepsilon^{n+1}$, which
generates a new "trajectory",  $|\varphi(t)\rangle^{n+1}$, according to the
Schr\"odinger equation.
We again relax the maximum to an extremum condition,
\begin{equation}
  \label{eq:extremR2}
  \frac{\partial R^{(1)}}{\partial \varepsilon}
  \bigg|_{\varepsilon^{n+1},|\varphi(t)\rangle^{n+1}} = 0\,,
\end{equation}
which gives
\begin{eqnarray*}
  0 &=& \frac{\partial}{\partial \varepsilon} \left[
    -g(\varepsilon) + \frac{\partial \Phi^{(1)}}{\partial t} +
    \mathfrak{Re}\left\langle\frac{\partial \Phi^{(1)}}{\partial \varphi} \bigg| 
      f(t,\varphi,\epsilon)\right\rangle
  \right]_{\varepsilon^{n+1},|\varphi(t)\rangle^{n+1}} \\
    &=& -\frac{\partial g(\varepsilon)}{\partial \varepsilon}
    \bigg|_{\varepsilon^{n+1}} + 
    \frac{1}{2}\left\langle\gamma(t) 
    \bigg| \frac{\partial f(t,\varphi,\epsilon) }{\partial \varepsilon}
    \right\rangle_{\varepsilon^{n+1},|\varphi\rangle^{n+1}} +
    \frac{1}{2} \left\langle 
      \frac{\partial f(t,\varphi,\epsilon) }{\partial \varepsilon}
    \bigg| \gamma(t)\right\rangle  _{\varepsilon^{n+1},|\varphi\rangle^{n+1}}
\end{eqnarray*}
and therefore
\begin{equation}
  \label{eq:condg}
  \frac{\partial g(\varepsilon)}{\partial \varepsilon}\bigg|_{\varepsilon^{n+1}} =
  \mathfrak{Re}\left\langle\gamma(t)\bigg|
   \frac{\partial f(t,\varphi,\epsilon) }{\partial \varepsilon}
  \right\rangle_{\varepsilon^{n+1},|\varphi\rangle^{n+1}} \,.
\end{equation}
With our choice of the constraint $g(\varepsilon)$, Eq.~(\ref{eq:g}), and 
using $\Op{H}=\Op{H}_0+\Op{\mu}\varepsilon(t)$ in the calculation
of $|\partial f / \partial \varepsilon\rangle$ , we obtain for the new
field
\begin{equation}
  \label{eq:neweps}
  \varepsilon^{n+1}(t) = \varepsilon^n(t) + 
  \frac{S(t)}{2\alpha}  \mathfrak{Im}\left[
     \langle \gamma(t) | \Op{\mu} |\varphi(t)\rangle^{n+1}
    \right] \,.
\end{equation}
The time evolution of the new state is given
 $|\varphi(t)\rangle^{n+1} =  \Op{U}(t,0;\varepsilon^{n+1}) |\varphi_i \rangle$, 
and $|\gamma(t)\rangle$ is obtained by formally solving Eq.~(\ref{eq:EoMaux}),
$|\gamma(t)\rangle = \Op{U}(t,T;\varepsilon^n)|\gamma(T)\rangle$,
\begin{displaymath}
  \varepsilon^{n+1}(t) = \varepsilon^n(t) + 
  \frac{S(t)}{2\alpha}  \mathfrak{Im}\left[
     \langle \gamma(T) |\Op{U}^+(t,T;\varepsilon^n) \Op{\mu} 
     \Op{U}(t,0;\varepsilon^{n+1})|\varphi_i\rangle
    \right] \,.  
\end{displaymath}
Using
the definition of the objective $F$, Eq.~(\ref{eq:goal}),
 $|\gamma(T)\rangle$ is determined
according to Eq.~(\ref{eq:initaux}),
$  |\gamma(T)\rangle = c  |\varphi_f\rangle $
with the coefficient 
\begin{equation}
  \label{eq:coef}
  c= \langle \varphi_f|\varphi(T)\rangle^n   =
  \langle  \varphi_f|\Op{U}(T,0;\varepsilon^n) |\varphi_i\rangle  
\end{equation}
and finally Eq.~({\ref{eq:deltaeps}}) is obtained.

Since we relaxed the maximum and mininum conditions to extremum conditions, we
need to ensure that each iteration indeed improves the objective, 
$J[\varphi^n,\varepsilon^n] -J[\varphi^{n+1},\varepsilon^{n+1}] \ge 0$,
\begin{eqnarray*}
  J[\varphi^n,\varepsilon^n] -J[\varphi^{n+1},\varepsilon^{n+1}] &=&
  L[\varphi^n,\varepsilon^n;\Phi^{(1)}] -
  L[\varphi^{n+1},\varepsilon^{n+1};\Phi^{(1)}] \\
  &=& \Delta_1 + \int_0^T \Delta_2(t) \Fkt{d}t
\end{eqnarray*}
with
\begin{eqnarray*}
  \Delta_1 &=& G^{(1)}[\varphi(T)^n] -G^{(1)}[\varphi(T)^{n+1}] \,,\\
  \Delta_2(t) &=&  R^{(1)}[t,\varphi(t)^{n+1},\varepsilon^{n+1}] - 
  R^{(1)}[t,\varphi(t)^n,\varepsilon^n] \,,
\end{eqnarray*}
A sufficient condition for 
$J[\varphi^n,\varepsilon^n] -J[\varphi^{n+1},\varepsilon^{n+1}] \ge 0$ is
$\Delta_1,\Delta_2(t)\ge 0$. 
Constructing $G^{(1)}$ from Eq.~(\ref{eq:Phi}), 
and using $  |\gamma(T)\rangle = c  |\varphi_f\rangle $ with Eq.~(\ref{eq:coef})
we see that
\begin{equation}
  \label{eq:cond1}
  \Delta_1 = \left|\langle \varphi_i |\Op{U}^+(T,0;\eps^n) -\Op{U}^+(T,0;\eps^{n+1})|
  \varphi_f\rangle\right|^2 \ge 0 \,.
\end{equation}
Because of the linearity of $f(t,\varphi,\varepsilon)$ 
w.r.t. $|\varphi\rangle$,
$R^{(1)}(t,\varphi,\epsilon^n)=-g(\varepsilon^n)$ for any $|\varphi\rangle$,
and we find
\begin{displaymath}
  \Delta_2(t) = R^{(1)}[t,\varphi(t)^{n+1},\varepsilon^{n+1}] - 
  R^{(1)}[t,\varphi(t)^{n+1},\varepsilon^n] \,.
\end{displaymath}
Constructing $R^{(1)}$ from Eq.~(\ref{eq:Phi}) and using 
again the linearity of the equation of motion,
we obtain as condition for monotonous convergence
\begin{equation}
  \label{eq:cond2}
  \Delta_2(t) = -g(\varepsilon^1) + g(\varepsilon^0) + 
  \left( \varepsilon^1 - \varepsilon^0 \right) 
  \frac{\partial g}{\partial \varepsilon}\Big\vert_{\varepsilon^1} \geq 0 \,.
\end{equation}

Eq.~(\ref{eq:condg}) and Eq.~(\ref{eq:cond2}) are the central equations
of the algorithm.
The constraints which we formulate in $g(\varepsilon)$ must fulfill 
Eq.~(\ref{eq:cond2}) to ensure convergence of the algorithm. 
Throughout this work we have used the constraint of restricted change in pulse energy,
i.e. $g(\varepsilon)$ as given by Eq.~(\ref{eq:g}),
which respects Eq.~(\ref{eq:cond2}) for all $\alpha$ as one can easily verify.

\section{Restriction of the spectral bandwidth}
\label{sec:restrict}
We tried other functional forms of the constraint $g(\varepsilon)$, in particular
\begin{equation}
  \label{eq:g_rest}
  g(\varepsilon) = \alpha_1(t) \left(\varepsilon - \varepsilon^0\right)^2 -
  \alpha_2(t) \left(\varepsilon - \varepsilon^\mathrm{ref}\right)^2
\end{equation}
with $\alpha_i= \alpha_{0i}/S(t)$,
where the second term restricts the new field
to a reference field $\varepsilon^\mathrm{ref}$ with a prespecified bandwidth
(note, that the negative sign of this term implies maximization, i.e. the
new field should be as close as possible to the reference field).
Condition~(\ref{eq:cond2}) is fulfilled for all
$\alpha_{01} \geq\alpha_{02}\geq 0$. We found no effect for  
$\alpha_{02} \ll \alpha_{01}$ as not enough weight is put on the second
constraint. Still 96\% of the objective can be reached
for $\alpha_{02}$ close to $\alpha_{01}$, but the spectral
bandwidth is slightly reduced. In particular, spurious high-frequency
components in the optimal puls can be avoided. The overall restriction
of the bandwidth, however, appears to be insufficient. 

In fact, it is not surprising that a condition formulated in time domain
as Eq.~(\ref{eq:g_rest}) cannot achieve the desired control over a frequency
domain property. This is a general problem of {\em global} (in time) 
optimal control schemes. Attempts to restrict the spectral bandwidth of the
puls have been made before~\cite{GrossJCP92,HornungJCP01}.
However, this is possible only at the cost of monotonic convergence.
An approach which unifies global optimal control with constraints in 
frequency domain will have to treat time and frequency on the same footing.
This shall be the subject of future work.


\end{document}